\documentclass[aps,prd,twocolumn,showpacs,superscriptaddress,nofootinbib,preprintnumbers]{revtex4-2}
\pdfoutput=1

\usepackage{graphicx}
\usepackage{bm}
\usepackage{amssymb,amsmath,latexsym}
\usepackage{xcolor}
\usepackage{hyperref}
\usepackage{cancel}
\usepackage{soul}
\usepackage{cancel}
\usepackage{placeins}

\def\ba{\begin{eqnarray}}
\def\ea{\end{eqnarray}}
\def\be{\begin{equation}}
\def\ee{\end{equation}}

\def\qtraj{\texttt{QTraj} }

\begin{document}

\title{Bottomonium production in heavy-ion collisions using quantum trajectories: Differential observables and momentum anisotropy}
\preprint{TUM-EFT 147/21; HU-EP-21/18-RTG}

\author{Nora Brambilla}
\email{nora.brambilla@ph.tum.de}
\affiliation{Physik-Department, Technische Universit\"{a}t M\"{u}nchen, James-Franck-Str. 1, 85748 Garching,
	Germany}
\affiliation{Institute for Advanced Study, Technische Universit\"{a}t M\"{u}nchen, Lichtenbergstrasse 2 a, 85748
	Garching, Germany}
\affiliation{Munich Data Science Institute, Technische Universit\"{a}t M\"{u}nchen, Walther-von-Dyck-Strasse 10, 85748 Garching, Germany}

\author{Miguel \'{A}ngel Escobedo}
\email{miguelangel.escobedo@usc.es}
\affiliation{Instituto Galego de F\'{i}sica de Altas Enerx\'{i}as (IGFAE), Universidade de Santiago de Compostela. E-15782, Galicia, Spain}

\author{Michael Strickland}
\email{mstrick6@kent.edu}
\affiliation{Department of Physics, Kent State University, Kent, OH 44242, United States}

\author{Antonio Vairo}
\email{antonio.vairo@tum.de}
\affiliation{Physik-Department, Technische Universit\"{a}t M\"{u}nchen, James-Franck-Str. 1, 85748 Garching, Germany}

\author{Peter Vander Griend}
\email{vandergriend@tum.de}
\affiliation{Physik-Department, Technische Universit\"{a}t M\"{u}nchen, James-Franck-Str. 1, 85748 Garching, Germany}

\author{Johannes Heinrich Weber}
\email{johannes.weber@physik.hu-berlin.de}
\affiliation{Department of Computational Mathematics, Science and Engineering, and Department of Physics and Astronomy, Michigan State University, East Lansing, MI 48824, USA}
\affiliation{Institut f\"ur Physik, Humboldt-Universit\"at zu Berlin \& IRIS Adlershof, D-12489 Berlin, Germany}

\begin{abstract}
	We report predictions for the suppression and elliptic flow of the $\Upsilon(1S)$, $\Upsilon(2S)$, and $\Upsilon(3S)$ as a function of centrality and transverse momentum in ultra-relativistic heavy-ion collisions.  
	We obtain our predictions by numerically solving a Lindblad equation for the evolution of the heavy-quarkonium reduced density matrix derived using potential nonrelativistic QCD and the formalism of open quantum systems.  
	To numerically solve the Lindblad equation, we make use of a stochastic unraveling called the quantum trajectories algorithm.  
	This unraveling allows us to solve the Lindblad evolution equation efficiently on large lattices with no angular momentum cutoff.  
	The resulting evolution describes the full 3D quantum and non-abelian evolution of the reduced density matrix for bottomonium states.  
	We expand upon our previous work by treating differential observables and elliptic flow; this is made possible by a newly implemented Monte-Carlo sampling of physical trajectories.
	Our final results are compared to experimental data collected in $\sqrt{s_{NN}} = 5.02$ TeV Pb-Pb collisions by the ALICE, ATLAS, and CMS collaborations.
\end{abstract}

\date{\today}

\keywords{Bottomonium suppression, Effective field theory methods, Open quantum system methods, Quark-gluon plasma, Relativistic heavy-ion collisions, Quantum chromodynamics}

\maketitle

\section{Introduction} \label{sec:intro}

Ultra-relativistic nucleus-nucleus (AA) collisions performed at the Relativistic Heavy Ion Collider (RHIC) at Brookhaven National Laboratory and the Large Hadron Collider (LHC) at the European Organization for Nuclear Research (CERN) have provided unprecedented insight into the behavior of matter at extreme energy and baryon number densities the likes of which previously only existed in the very early Universe \cite{Averbeck2015,Busza:2018}.  
The goal of these experiments is to produce and study a color-ionized, or deconfined, quark-gluon plasma (QGP), a state of matter in which the degrees of freedom are quarks and gluons rather than the hadronic degrees of freedom observed at low energies in which quarks and gluons are confined.  
In order to determine the properties of the QGP, experimentalists at RHIC and LHC measure a variety of observables in AA, pA, and  pp collisions including the spectra of produced hadrons, their azimuthal momentum correlations, photon production, dilepton production, etc.  

An observable of particular interest is the ratio of the number of heavy quarkonia observed in an AA collision to the number observed in a pp collision (scaled by the number of binary collisions); this defines the nuclear modification factor $R_{AA}$ of the particular quarkonium species.
It was predicted decades ago that due to Debye screening at distances larger than approximately the inverse of the Debye mass, the inter-quark potential of heavy quarkonium in a color-ionized QGP becomes more short range,  and consequently, the measured rates of heavy quarkonium bound state production in AA collisions would be suppressed relative to the rates in pp collisions in which no QGP is generated \cite{Matsui:1986dk,Karsch:1987pv}.  
Since these early papers, there has been considerable progress in understanding the dynamics of heavy quarkonia in the QGP.  
A paradigmatic shift in our theoretical understanding of heavy quarkonium suppression occurred in 2007 with the findings of thermal corrections to the real part of the in-medium potential related to screening and a nonzero imaginary part related to the in-medium dissociation rate due to Landau damping \cite{Laine:2006ns}.
Subsequent works extended this to include the effect of non-abelian singlet-octet transitions using the effective field theory (EFT) potential non-relativistic QCD (pNRQCD)~\cite{Brambilla:2008cx,Escobedo:2008sy,Brambilla:2010vq,Beraudo:2007ky}.  
In the interim, the existence of a large in-medium decay width has been taken into account in phenomenological calculations of $R_{AA}$ which use complex potential models~\cite{Strickland:2011mw,Strickland:2011aa,Krouppa:2015yoa,Krouppa:2016jcl,Krouppa:2017jlg,Islam:2020gdv,Islam:2020bnp}.
Nonrelativistic EFTs, and especially pNRQCD, allow for a systematic and non-perturbative exploitation of the separation of scales inherent in heavy quark bound states.

In order to fully understand the dynamics of in-medium heavy quarkonium, a careful consideration of in-medium scattering including both dissociation and recombination is necessary. 
The formalism of open quantum systems (OQS) allows for a rigorous treatment of a quantum system (here the heavy quarkonium) coupled to an external environment (here the QGP) and thus provides a useful framework for treating heavy quarkonia in medium~\cite{Akamatsu:2014qsa,Rothkopf:2019ipj,Akamatsu:2020ypb,Yao:2021lus}.
In the present work, we utilize a set of evolution equations describing the in-medium evolution of heavy quarkonium realizing the hierarchy of scales $1/a_{0} \gg \pi T \sim m_D \gg E$ where $a_{0}$ is the Bohr radius of the bound state, $T$ is the medium temperature, $m_{D}\sim gT$ is the Debye screening mass, and $E$ is the binding energy of the bound state.
In this regime, the evolution equations take the form of a Lindblad equation describing the Markovian quantum Brownian motion of a heavy quarkonium in the QGP~\cite{Brambilla:2016wgg,Brambilla:2017zei,Brambilla:2019tpt}.

In this work, we extend Ref.~\cite{Brambilla:2020qwo} wherein the Lindblad equation was solved numerically using the quantum trajectories algorithm which represents a quantum unraveling of the Lindblad equation.
The numerical code, developed for and presented in Ref.~\cite{Brambilla:2020qwo}, is called \qtraj and was used to make phenomenological predictions for the nuclear suppression of $\Upsilon(1S)$, $\Upsilon(2S)$, and $\Upsilon(3S)$ states in 5.02 TeV Pb-Pb collisions as a function of the number of participating nucleons $N_\text{part}$.  
The quantum trajectories algorithm requires averaging over a set of stochastically-generated quantum evolutions.   
Due to the associated computational costs, in Ref.~\cite{Brambilla:2020qwo}, the temperature evolution of the plasma was simplified by using an average temperature profile per centrality class computed from the average of Monte-Carlo sampled physical trajectories in that centrality class. 
In this work, we compute the QGP survival probability for each physical trajectory and bin the results as is done experimentally.  
This has been made possible by efficiency and scalability improvements to the \qtraj code~\cite{compforth}.
As a result of these improvements, we are able to present predictions for $R_{AA}$ and associated double ratios as functions of both $N_\text{part}$ and $p_T$.  
In addition, due to the large number of physical trajectories now considered, we are able to make statistically significant predictions for the elliptic flow $v_{2}$ of the $\Upsilon(1S)$, $\Upsilon(2S)$, and $\Upsilon(3S)$ states as functions of both $N_\text{part}$ and $p_T$.  
We compare our results to experimental data collected by the ALICE, ATLAS, and CMS collaborations.

The structure of this work is as follows: 
in Sec.~\ref{sec:methodology}, we review the derivation of the Lindblad equation describing in-medium heavy-quarkonium dynamics in a strongly-coupled QGP and the quantum trajectories algorithm as implemented in the \qtraj code;
in Sec.~\ref{sec:results}, we present our numerical results and compare to experimental data;
in Sec.~\ref{sec:conclusions}, we present our conclusions and an outlook for the future;
in App.~\ref{app:table}, we present a table of \qtraj predictions for the centrality-integrated $R_{AA}$ of the $\Upsilon(1S)$, $\Upsilon(2S)$, and $\Upsilon(3S)$;
finally, in App.~\ref{app:jump_nojump}, we investigate the role of quantum jumps in heavy-quarkonium dynamics and their effect on experimental observables.

\section{Methodology} \label{sec:methodology}

\subsection{Heavy quarkonium dynamics in a strongly-coupled quark-gluon plasma} \label{subsec:theory}

In this paper, we solve the Lindblad equation describing the in-medium dynamics of a heavy quarkonium
that was derived using the EFT pNRQCD and the OQS formalism in Refs.~\cite{Brambilla:2016wgg,Brambilla:2017zei,Brambilla:2019tpt}.
The nonrelativistic nature of heavy-heavy bound states, i.e., $v\ll 1$ where $v$ is the quark-antiquark relative velocity, leads to at least three hierarchically ordered scales:
the hard scale $M$ of the heavy quark mass, the soft scale $Mv$ of typical momentum transfers, and the ultrasoft scale $Mv^{2}$ associated with the binding energy $E$.
If the bound state is Coulombic then $v\sim\alpha_{s}$.
Integrating out the hard scale $M$ from full QCD gives rise to the EFT nonrelativistic QCD (NRQCD) \cite{Caswell:1985ui,Bodwin:1994jh};
further integrating out the soft scale $Mv$ gives rise to pNRQCD \cite{Pineda:1997bj,Brambilla:1999xf,Brambilla:2004jw}.
In this treatment, the small radius $r$ of the lowest lying bound states allows for a multipole expansion in $r$.
pNRQCD implements this expansion in the bound state radius $r$ and in the inverse of the heavy quark mass $M$ at the Lagrangian level
and is thus ideally suited for describing low lying bottomonium states of small radius.
The degrees of freedom in the resulting effective Lagrangian are composite fields made of heavy quark and heavy antiquark pairs in a color singlet or color octet configuration, 
and light quarks and gluons at the ultrasoft scale.
Transitions between the singlet and octet fields are encoded in chromoelectric-dipole interaction terms.  

The OQS formalism allows for the rigorous treatment of a quantum system coupled to an external environment (see Ref.~\cite{Breuer:2002pc} for a general introduction).
The relevant time scales of the full system are 
a time scale $\tau_{S}$ characterizing the system, a time scale $\tau_{E}$ characterizing the environment,
and a relaxation time $\tau_{R}$ characterizing the interaction between the system and the environment.
The scale $\tau_{S}$ is set by the characteristic time scale of internal transitions in the system and, as such, is related to the inverse of the internal level spacing of states.  
The scale $\tau_{E}$ is set by the time scale of equilibration of the environment, and the scale  $\tau_{R}$ is the characteristic time scale associated with the in-medium evolution of the reduced density matrix.
Hierarchical orderings of these scales allow for simplifications of calculations and the realization of different evolution paradigms.
For the system treated in this work, i.e., a bottomonium in a QGP at temperatures reached in current heavy ion collision experiments, one has
\be\label{eq:markov}
\tau_{R} \gg \tau_{E} \, ,
\ee
which allows for the Markovian approximation, i.e., the system is insensitive to its prior evolution.
Furthermore, one has
\be\label{eq:qbm}
\tau_{S} \gg \tau_{E} \, ,
\ee 
which qualifies the evolution as quantum Brownian motion.

We consider a strongly coupled plasma in which the heavy-quark mass $M$, the Bohr radius of the quarkonium $a_{0}$, the temperature of the medium $T$,
the Debye mass $m_{D} \sim gT$, and the binding energy of the quarkonium $E$ fulfill the hierarchy of scales
\begin{equation}\label{eq:hierarchy}
M \gtrsim 1/a_{0} \gg \pi T \sim m_{D} \gg E \, .
\end{equation}
In this regime, the system, the environment, and the relaxation time scales are given by
\begin{eqnarray}
\tau_{S} &\sim& \frac{1}{E} \, , \\
\tau_{E} &\sim& \frac{1}{\pi T} \, , \\
\tau_{R} &\sim& \frac{1}{\Sigma_{s}} \sim \frac{1}{a_{0}^{2} (\pi T)^{3}} \, ,
\end{eqnarray}
where $\Sigma_{s}$ is the thermal self-energy of the system.
The hierarchy of scales in Eq.~(\ref{eq:hierarchy}) ensures that the evolution of the reduced density matrix is Markovian and exhibits quantum Brownian motion.

Using pNRQCD and OQS and working in the regime specified in Eq.~(\ref{eq:hierarchy}),
in Refs. \cite{Brambilla:2016wgg,Brambilla:2017zei} a set of master equations governing the in-medium evolution of a heavy quarkonium was derived.
In the limit $T \gg E$, an expansion in $E/T$ may be performed; at leading order, the evolution equations take the form of a Lindblad equation \cite{Lindblad:1975ef,Gorini:1975nb}
\begin{equation}\label{eq:lindblad}
\frac{d \rho(t)}{dt} = -i[H, \rho(t)] + \sum_{n} \left( C_{n} \rho(t) C_{n}^{\dagger} - \frac{1}{2} \left\{ C_{n}^{\dagger}C_{n}, \rho(t) \right\} \right),
\end{equation}
where
\begin{align}
\rho(t) =& \begin{pmatrix} \rho_{s}(t) & 0 \\ 0 & \rho_{o}(t) \end{pmatrix},\\
\label{eq:h_with_gamma}
H =& \begin{pmatrix} h_{s} & 0 \\ 0 & h_{o} \end{pmatrix} + \frac{r^{2}}{2} \gamma \begin{pmatrix} 1 & 0 \\ 0 & \frac{N_{c}^{2}-2}{2(N_{c}^{2}-1)} \end{pmatrix},\\
\label{eq:c0}
C_{i}^{0} =& \sqrt{\frac{\kappa}{N_{c}^{2}-1}} r^{i} \begin{pmatrix} 0 & 1 \\ \sqrt{N_{c}^{2}-1} & 0 \end{pmatrix},\\
\label{eq:c1}
C_{i}^{1} =& \sqrt{\frac{(N_{c}^{2}-4)\kappa}{2(N_{c}^{2}-1)}} r^{i} \begin{pmatrix} 0 & 0 \\ 0 & 1 \end{pmatrix}.
\end{align}
The singlet and octet density matrices $\rho_{s}(t)$ and $\rho_{o}(t)$ describe quarkonium in the singlet and octet configurations, respectively.
The operators $h_{s,o}={\mathbf{p}^{2}}/{M} + V_{s,o}$ are the singlet and octet Hamiltonians with $V_{s}=-4\alpha_{s}(1/a_{0})/(3r)$ and $V_{o}=\alpha_{s}(1/a_{0})/(6r)$;
$\alpha_{s}(1/a_{0})$ is the strong coupling at the energy scale of the inverse of the Bohr radius.
Interactions with the strongly-coupled medium are encoded in the non-perturbative transport coefficients $\kappa$ and $\gamma$
\ba
\kappa &=& \frac{g^{2}}{18} \int_{0}^{\infty} dt \left\langle \left\{ \tilde{E}^{a,i}(t,\mathbf{0}), \tilde{E}^{a,i}(0,\mathbf{0}) \right\} \right\rangle, \label{eq:kappadef} \\
\gamma &=& -i \frac{g^{2}}{18} \int_{0}^{\infty} dt \left\langle \left[ \tilde{E}^{a,i}(t,\mathbf{0}), \tilde{E}^{a,i}(0,\mathbf{0}) \right] \right\rangle  \label{eq:gammadef},
\ea
where 
\be
\tilde{E}^{a,i}(t,\mathbf{0}) = \Omega^{\dagger}(t) E^{a,i}(t,\mathbf{0}) \Omega(t) \, ,
\ee
with $\Omega(t)$ being a temporal Wilson line running from time negative infinity to time $t$, i.e.,
\be
\Omega(t) = \text{exp}\left[ -i g \int_{-\infty}^{t} dt' A_{0}(t', \textbf{0}) \right].
\ee
$\kappa$ is the heavy quark momentum diffusion coefficient \cite{CasalderreySolana:2006rq,CaronHuot:2007gq}, and $\gamma$ is its dispersive counterpart.
As noted in Ref.~\cite{Brambilla:2017zei}, $\kappa$ and $\gamma$ are related to the thermal width $\Gamma$ and mass shift $\delta M$ of the bottomonium, respectively,
and can, therefore, be extracted indirectly from unquenched lattice measurements of these quantities as done in Ref.~\cite{Brambilla:2019tpt}.
More recently, direct quenched lattice measurements of $\kappa$ have been performed across an unprecedentedly large range of temperatures
allowing to detect the dependence of $\kappa$ on the medium temperature \cite{Brambilla:2020siz}.
Direct lattice extractions of $\gamma$ (as opposed to the indirect extractions via $\delta M$ in Ref.~\cite{Brambilla:2019tpt}) are currently in progress.

\subsection{Quantum trajectories algorithm} \label{subsec:qtraj}

Directly solving the Lindblad equation given in Sec.~\ref{subsec:theory} is computationally demanding, and previous works relied on simplifying assumptions.
Specifically, Ref.~\cite{Brambilla:2017zei} expanded the density matrix in spherical harmonics and introduced a cutoff at $\ell=1$, thus only considering $S$- and $P$-wave states.
In Ref.~\cite{Brambilla:2020qwo}, the quantum trajectories algorithm was utilized to solve the Lindblad equation via a computationally less intensive Monte-Carlo method.
This allowed for solving of the evolution equations to all orders in $\ell$ while also dramatically increasing the spatial extent of the lattice and decreasing the lattice spacing compared to Ref.~\cite{Brambilla:2017zei}.

The quantum trajectories algorithm implements a stochastic evolution of each quantum trajectory in order to solve the Lindblad equation (frequently referred to as an unraveling of the Lindblad equation).\footnote{For a comprehensive introduction to this method see Ref.~\cite{Daley:2014fha}.}
The central idea of the algorithm is to split the full evolution specified by the Lindblad equation into a diagonal contribution that leaves the quantum numbers of the system unchanged and an off-diagonal contribution that changes the quantum numbers.
For this purpose, we rewrite the Lindblad equation as
\begin{equation}\label{eq:heff_lindblad}
\frac{d \rho(t)}{dt} = -iH{_\text{eff}} \rho(t) + i \rho(t) H^{\dagger}_{\text{eff}} + \sum_{n} C_{n} \rho(t) C_{n}^{\dagger} \, ,
\end{equation}
where 
\begin{equation}
H_\text{eff} = H - \frac{i}{2}\sum_{n} C_{n}^{\dagger}C_{n}.
\end{equation}
The non-unitary effective Hamiltonian $H_\text{eff} $ is diagonal; its action on $\rho(t)$ leaves the color and angular momentum state of $\rho(t)$ unchanged but decreases its trace.
The jump operators $C_{n}$ entering into the summation in Eq.~\eqref{eq:heff_lindblad}  are off-diagonal and their action on $\rho(t)$ results in a change of quantum numbers.\footnote{
	This is clearly the case for $C_{0}$ as it is off diagonal in color space, i.e., it induces a singlet-octet transition (and a change of $\pm1$ in $\ell$).  
	$C_{1}$ is diagonal in color space but off diagonal in angular momentum space, i.e., it induces an octet-octet transition between states of angular momentum $\ell$ and $\ell\pm1$.
	This can be made manifest by expanding in spherical harmonics; cf. Eqs.~(83) and (84) of \cite{Brambilla:2017zei}.}
The diagonal contributions include the effect of the thermal width $\Gamma=\sum_{n} C_{n}^{\dagger}C_{n}$ in the evolution (cf. Eq.~(2.2) of Ref.~\cite{Brambilla:2020qwo}), and the off-diagonal terms can be mapped to \textit{quantum jumps} between different states.
Both of these contributions can be implemented at the level of one-dimensional wave functions rather than density matrices, thereby greatly reducing both the memory needed for the simulation and the number of computational cycles required.\footnote{  
	Details concerning the \qtraj implementation, including scaling studies, benchmarks, and runtime comparisons to other methods can be found in Ref.~\cite{compforth}.  This reference accompanies the open-source release of \qtraj\!\!.
}

The \qtraj code implements the quantum trajectories algorithm as follows:
\begin{enumerate}
	\item Initialize a wave function $|\psi(t_{0})\rangle$ at initial time $t_{0}$ which corresponds to the initial quantum state of the particle given by $\rho(t_{0})= |\psi(t_{0})\rangle\langle\psi(t_{0})|$.
	\item Generate a random number $0<r_{1}<1$ and evolve the wave function forward in time with $H_\text{eff} $ until
	\begin{equation}\label{eq:heff_evolution}
	|| \, e^{-i \int_{t_{0}}^{t}dt' H_\text{eff} (t')} | \psi(t_{0}) \rangle \, ||^{2} \le r_{1} \, .
	\end{equation}
	Denote the first time step fulfilling the inequality of Eq.~(\ref{eq:heff_evolution}) as the jump time $t_{j}$.  If the jump time is greater than the simulation run time $t_{f}$, end the simulation at time $t_{f}$; otherwise, proceed to step~\ref{step:jump}.
	\label{step:evolution}	
	\item At time $t_{j}$, initiate a quantum jump:
	\begin{enumerate}
		\item If the system is in a singlet configuration, jump to octet.  If the system is in an octet configuration, generate a random number $0 < r_{2} < 1$ and jump to singlet if $r_{2}<2/7$; otherwise, remain in the octet configuration.
		\item Generate a random number $0<r_{3}<1$; if $r_{3}<l/(2l+1)$, take $l\to l-1$; otherwise, take $l \to l+1$.
		\item Multiply the wavefunction by $r$ and normalize.
	\end{enumerate}
	\label{step:jump}
	\item Continue from step~\ref{step:evolution}.
\end{enumerate}
The procedure for the calculation of the jump time $t_{j}$ in step~\ref{step:evolution} is known as the \textit{waiting time approach} and reduces the number of random numbers to be generated compared to the standard quantum trajectories approach (see Sec.~III.D of Ref.~\cite{Daley:2014fha} and references therein).
The probabilities in step~\ref{step:jump} correspond to the branching fractions into a state of different angular momentum and/or color and are calculated via the relation
\begin{equation}
p_{n} = \frac{\langle \psi(t)| C_{n}^{\dagger} C_{n} | \psi(t) \rangle }{\sum_{n} \langle \psi(t)| C_{n}^{\dagger} C_{n} | \psi(t) \rangle}.
\end{equation}
Each evolution of the wave function from time $t_{0}$ to $t_{f}$ is called a \textit{quantum trajectory}.  
In practice, a large number of quantum trajectories must be generated and averaged over, and, as the number of trajectories considered increases, the average converges to the solution of the Lindblad equation.
This equivalence can be explicitly proven by writing $|\psi(t+\delta t)\rangle$ as a superposition of a jumped state and a state evolved with $H_\text{eff} $.  For details of this proof, see Sec.~III.A of Ref.~\cite{Daley:2014fha}.

\subsection{Simulation details} \label{subsec:simulation_details}

In order to solve Eq.~\eqref{eq:lindblad}, we must specify the values of the transport coefficients $\kappa$ and $\gamma$.
For the former, we make use of recent quenched lattice measurements of $\kappa$ carried out in Ref.~\cite{Brambilla:2020siz} which provide $\kappa(T)$ over a large range of temperatures.
All results reported in this work are carried out using three temperature-dependent parameterizations of $\hat{\kappa}(T)=\kappa(T)/T^{3}$ which are given by the lower, central, and upper bounds of the ``fit'' curve of Fig.~13 of \cite{Brambilla:2020siz}.  
We denote these three parameterizations $\hat{\kappa}_{L}(T)$, $\hat{\kappa}_{C}(T)$, and $\hat{\kappa}_{U}(T)$, respectively.
For $\gamma$, we perform simulations with three temperature-independent values of $\hat{\gamma} = \gamma / T^{3} = \{-3.5,\, -1.75, \, 0 \}$.
These values are taken from the relation $\delta M(1S) = (3/2)a_{0}^{2} \gamma$ where $\delta M(1S)$ is the in-medium mass shift of the $\Upsilon(1S)$ state as detailed in Ref.~\cite{Brambilla:2019tpt}.
We note that the lattice studies of Refs.~\cite{Kim:2018yhk,Aarts:2011sm} used in Ref.~\cite{Brambilla:2019tpt} favor larger absolute values of (the negative parameter) $\hat{\gamma}$, while more recent lattice studies \cite{Larsen:2019bwy,Shi:2021qri} favor $\delta M(\Upsilon(1S)) \simeq 0$ and thus $\hat{\gamma} \simeq 0$.

For the mass, we take $M = m_{b} = m_{\Upsilon(1S)}/2 = 4.73$ GeV with $m_{\Upsilon(1S)}$ from \cite{Zyla:2020zbs}.\footnote{
	We note that we update the value of $m_{b}$ used in this work compared to Refs.~\cite{Brambilla:2017zei,Brambilla:2020qwo} in order to be more consistent with other literature.  
	As a result, the value of $\alpha_{s}$ changes accordingly.
}
The strong coupling $\alpha_{s}$ is calculated by solving 
\be
a_{0} = \frac{2}{C_{F} \, \alpha_{s}(1/a_{0}) \, m_{b}},
\ee
where $\alpha_{s}$ is evaluated at the inverse of the Bohr radius using the 1-loop running with $N_{f}=3$ flavors, and $\Lambda_{\overline{MS}}^{N_f=3}=332$ MeV~\cite{Petreczky:2020tky}. 
The resulting value of the strong coupling constant is $\alpha_{s} = 0.468$.

For the initial state radial wave-function we use a Gaussian-smeared delta function multiplied by a power of $r$ appropriate for the initial angular momentum state $\ell$, i.e.,
\begin{equation}
\psi_{\ell}(t_0) \propto r^{\ell} e^{-r^{2}/(ca_{0})^2},
\end{equation} 
with $r\, \psi_{\ell}(t_{0})$ normalized to one when summed over the entire (one-dimensional) lattice volume. 
Narrower initial states (smaller $c$) require a significantly larger number of trajectories to obtain similar statistical errors. 
We take the width of the Gaussian to be $c = 0.2$ to balance accuracy and computational effort; while this choice may cause relative systematic uncertainties of about 10\% or 15\% for the excited $S$-wave states, the $S$-wave ground state is unaffected (below 5\% level) by changes of $c$ within a factor of two~\cite{compforth}.

We employ a radial lattice of $\texttt{NUM}=4096$ lattice sites and a radial volume of $\texttt{L}=80\,\mathrm{GeV}^{-1}$, corresponding to a radial lattice spacing of $a \approx 0.0195\,\mathrm{GeV}^{-1}$. 
Systematic errors due to the finite lattice spacing or volume are of the same order as those due to the smeared initial state;
the former is more significant for the ground state, the latter for the excited states. 
The real time integration employed for deterministic evolution between jumps is discretized with a time step of $\texttt{dt}=0.001\,\mathrm{GeV}^{-1}$; 
this time discretization leads to a quantitatively similar level of systematic errors as the other sources~\cite{compforth}.

We expand upon our work reported in Ref.~\cite{Brambilla:2020qwo} by Monte-Carlo generating independent physical trajectories through the quark-gluon plasma rather than using a single path-averaged temperature evolution in each centrality bin.
In Ref.~\cite{Brambilla:2020qwo}, in each centrality bin, a path-averaged temperature evolution was computed from the average of approximately 132000 Monte-Carlo generated physical trajectories and used to compute the survival probability.
In the present work, due to increased code efficiency/scalability and access to large-scale computational resources, we sample approximately 7\,-\,9\,$\times 10^5$ independent physical trajectories for each choice of $\hat\kappa(T)$ and $\hat\gamma$, with approximately 50-100 quantum trajectories per physical trajectory.  
To generate each physical trajectory, we sample the bottomonium production point in the transverse plane using the nuclear binary collision overlap profile $N^{\text{bin}}_{AA}(x,y,b)$, the initial transverse momentum of the state $p_T$ from an $E_T^{-4}$ spectrum, and the initial azimuthal angle $\phi$ of the state's momentum uniformly in $[0,2\pi)$.  
We bin the results for the survival probability as a function of centrality, $p_T$, and $\phi$.  
This allows us to make predictions for differential observables such as $R_{AA}$ as a function of $p_T$ and elliptic flow.

We use the same medium evolution as Ref.~\cite{Brambilla:2020qwo} that is modeled using a 3+1D dissipative relativistic hydrodynamics code, which makes use of the quasiparticle anisotropic hydrodynamics (aHydroQP) framework~\cite{Alqahtani:2015qja,Alqahtani:2016rth,Alqahtani:2017mhy}.
The code uses a realistic equation of state fit to lattice QCD measurements \cite{Bazavov:2013txa} and was tuned to soft hadronic data collected in 5.02 TeV collisions using smooth optical Glauber initial conditions in Ref.~\cite{Alqahtani:2020paa}.  
The resulting hydrodynamic parameters provide an excellent description of the experimentally observed hadronic spectra/multiplicities, extracted femtoscopic radii, and identified hadron elliptic flow with an initial central temperature of $T_0 = 630$ MeV at $\tau_0 = 0.25$ fm/c and a constant specific shear viscosity of $\eta/s = 0.159$.
The anisotropic hydrodynamics framework allows for an accurate description of both the early-time evolution of the quark-gluon plasma and the evolution near the transverse edges of the plasma where deviations from equilibrium are large.   
This is due to an all orders resummation in the inverse Reynolds number \cite{Strickland:2017kux}.
As a result, aHydroQP reliably describes even the very early stages of the collision, when non-equilibrium corrections are large, in addition to extreme cases of the flow profile, such as Gubser flow where non-equilibrium corrections are large both at early and late times~\cite{Gubser:2010ze,Gubser:2010ui,Nopoush:2014qba,Florkowski:2013lza,Florkowski:2013lya,Florkowski:2014sfa,Denicol:2014xca,Denicol:2014tha,Heller:2015dha,Keegan:2015avk,Strickland:2018ayk,Strickland:2019hff,Almaalol:2020rnu}.

In our simulations, the wave-function is initialized at time $\tau=0$ fm/c and evolved in the vacuum until the interaction with the medium is initialized at $\tau=0.6$ fm/c.
To ensure that the hierarchy of scales of Eq.~(\ref{eq:hierarchy}) is fulfilled and our evolution equations are valid, we evolve the state in the vacuum when the temperature falls below $T_{f}=250$ MeV.
In this temperature region, the hierarchy of scales given in Eq.~(\ref{eq:hierarchy}) is no longer fulfilled as $\pi T$ is no longer significantly greater than the binding energy $E$.
Hence, in this temperature region, the medium effects are ignored, and the quantum state is evolved using the vacuum potential.
As this particular value of $T_{f}$ is somewhat arbitrary, in Ref.~\cite{Brambilla:2020qwo}, a set of simulations were performed varying $T_{f}$ by $\pm25$ MeV; the uncertainty from this variation was found to be similar in magnitude to that obtained from variation of $\hat{\kappa}(T)$ and $\hat{\gamma}$.
We note that the most recent lattice quantum chromodynamics (LQCD) calculations find that the pseudocritical temperature for the QGP phase transition is approximately $T_{pc} \simeq 158\ \text{MeV}$ \cite{Bazavov:2018mes,Borsanyi:2020fev}.\footnote{A study is in progress to determine the next-to-leading-order corrections to the evolution equations in the $E/T$ expansion, thus extending the validity of the description to lower temperatures \cite{nlo}.}  
All results reported in this work are obtained using $T_{f}=250$ MeV.

\begin{figure*}[ht]
	\begin{center}
		\includegraphics[width=0.43\linewidth]{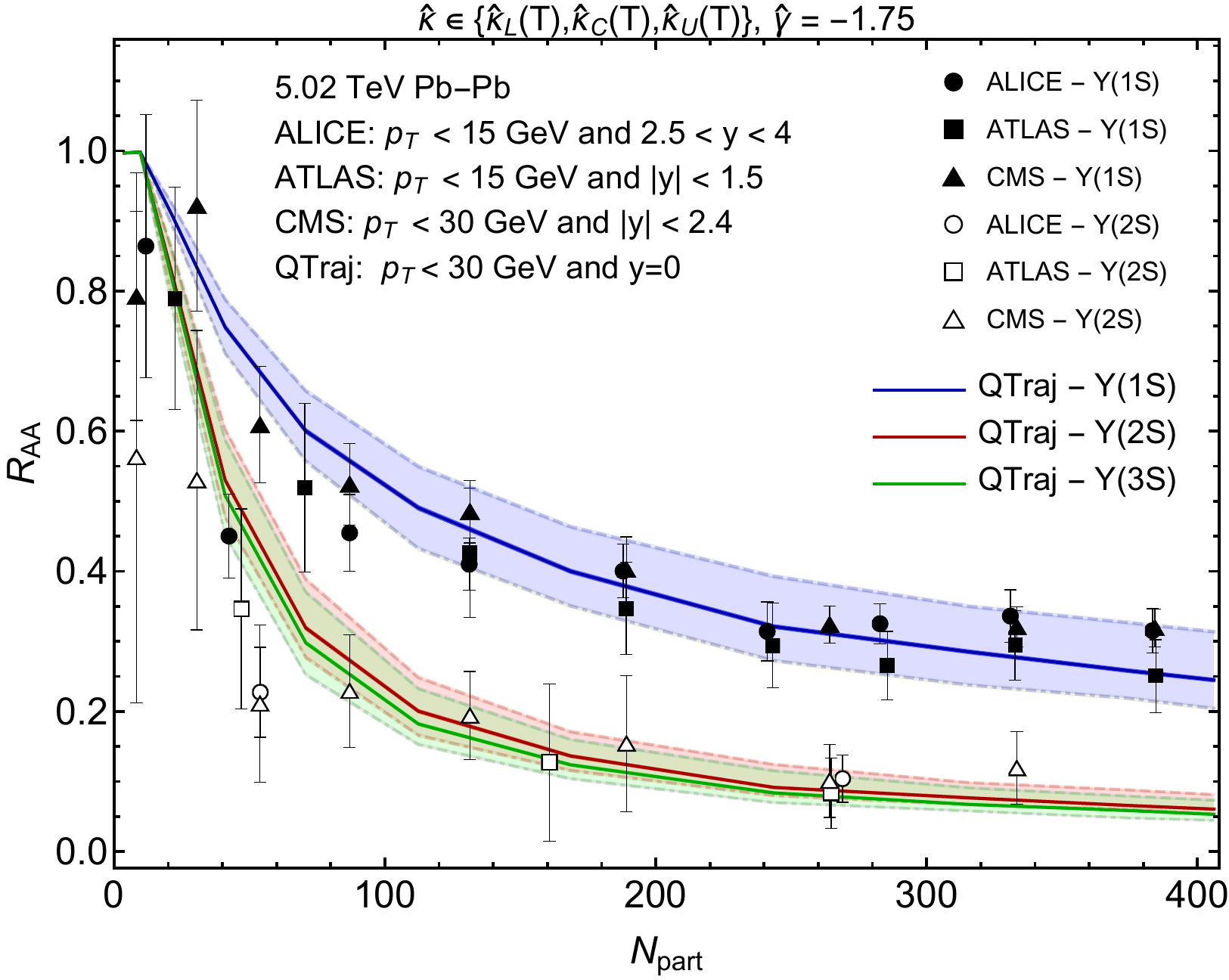}  \hspace{1cm}
		\includegraphics[width=0.43\linewidth]{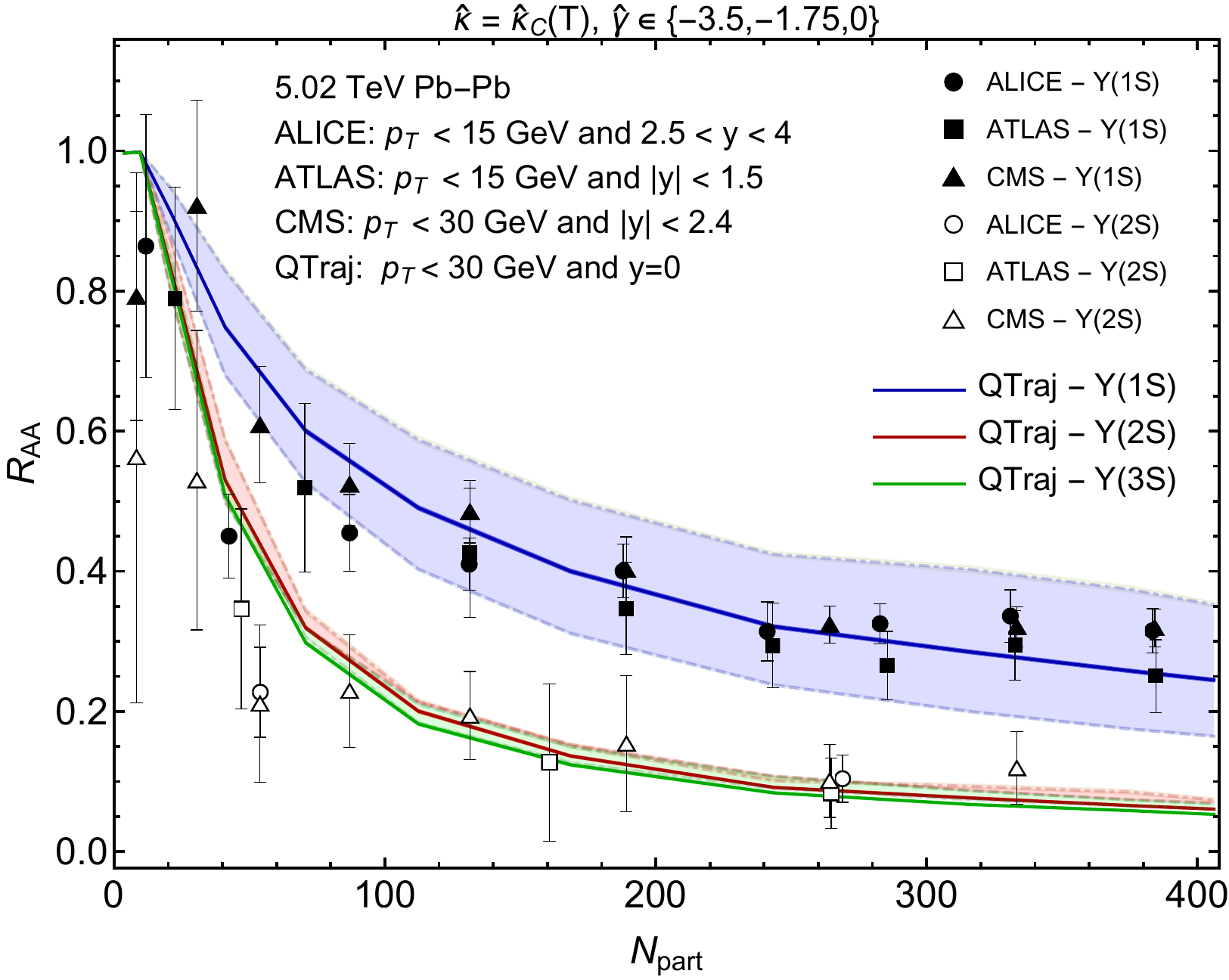}
	\end{center}
	\caption{(Color online)
		The nuclear modification factor $R_{AA}$ of the $\Upsilon(1S)$, $\Upsilon(2S)$, and $\Upsilon(3S)$ as a function of $N_{\text{part}}$ compared to experimental measurements from the ALICE~\cite{Acharya:2020kls}, ATLAS~\cite{ATLAS5TeV}, and CMS~\cite{Sirunyan:2018nsz} collaborations.
		The bands in the theoretical curves indicate variation with respect to $\hat{\kappa}(T)$ (left) and $\hat{\gamma}$ (right).
		The central curves represent the central values of $\hat{\kappa}(T)$ and $\hat{\gamma}$, and the dashed and dot-dashed lines represent the lower and upper values, respectively, of $\hat{\kappa}(T)$ and $\hat{\gamma}$. 
	}
	\label{fig:raa_vs_npart}
\end{figure*}

\begin{figure*}[ht]
	\begin{center}
		\includegraphics[width=0.43\linewidth]{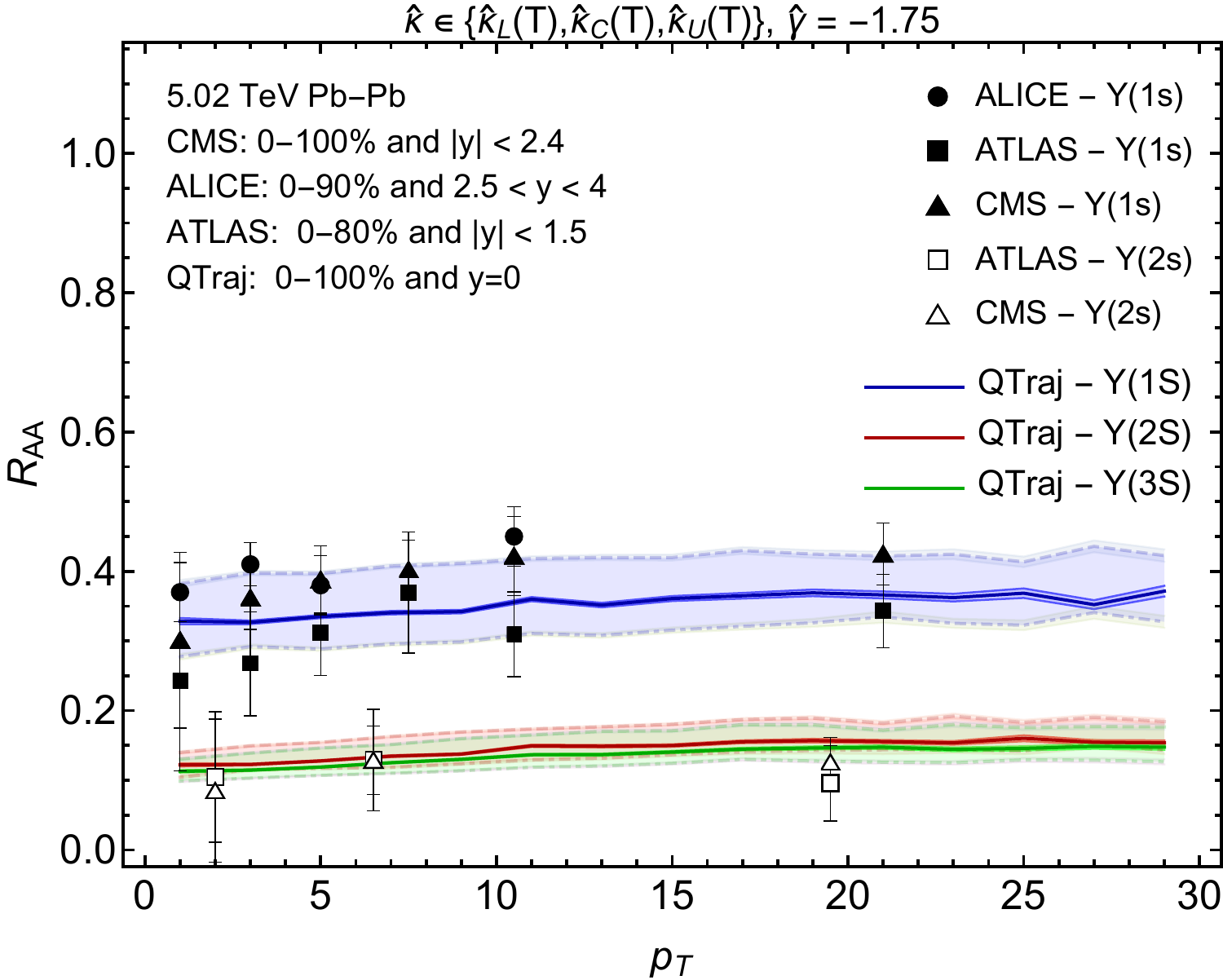}  \hspace{1cm}
		\includegraphics[width=0.43\linewidth]{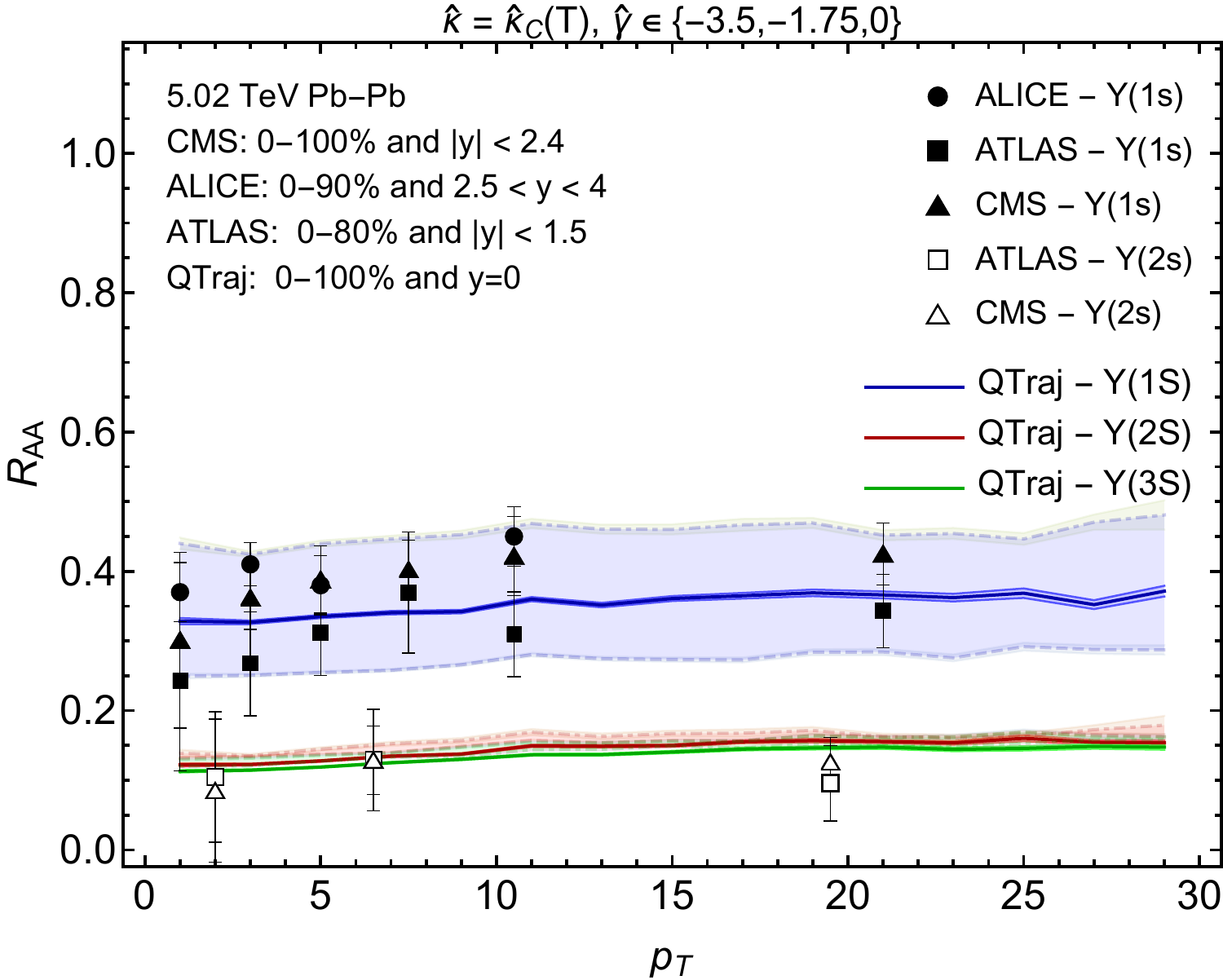}
	\end{center}
	\caption{(Color online)
		The nuclear modification factor $R_{AA}$ of the $\Upsilon(1S)$, $\Upsilon(2S)$, and $\Upsilon(3S)$ as a function of $p_{T}$ compared to experimental measurements.
		The experimental data are taken from the ALICE~\cite{Acharya:2020kls}, ATLAS~\cite{ATLAS5TeV}, and CMS~\cite{Sirunyan:2018nsz} collaborations.
		The bands represent theoretical uncertainties as in Fig.~\ref{fig:raa_vs_npart}.
	}
	\label{fig:raa_vs_pt}
\end{figure*}

\subsection{Feed down} \label{subsec:feeddown}

The \qtraj code allows for a computationally efficient solution of the Lindblad equation describing the in-medium evolution of bottomonium states in the QGP.   
From this evolution, one can extract the survival probability of a state that has traversed the QGP.
However, in order to compare to experimental measurements of the nuclear modification factor $R_{AA}$, one must take into account the probability that an excited bottomonium state emerging from the plasma decays to a lower-lying bottomonium state in the vacuum before being experimentally detected.
At the level of the cross section, the experimentally observed and direct production cross sections are related by $\vec{\sigma}_{\text{exp}} = F \vec{\sigma}_{\text{direct}}$ where each entry of the $\sigma$ vectors corresponds to a particular bottomonium state, and $F$ is a matrix related to the branching ratios of the excited states.
We consider the states $\{ \Upsilon(1S),\,$ $\Upsilon(2S),\,$ $\chi_{b0}(1P),\,$ $\chi_{b1}(1P),\,$ $\chi_{b2}(1P),\,$ $\Upsilon(3S),\,$ $\chi_{b0}(2P),\,$ $\chi_{b1}(2P),\,$ $\chi_{b2}(2P)\}$.
The entry $F_{ij}$ $i<j$ is the branching ratio of state $j$ to state $i$, $F_{ii}=1$, and $F_{ij}=0$ for $i>j$.  
The explicit values of $F_{ij}$ are taken from the Particle Data Group ~\cite{pdg} and presented in Eq.~(6.4) of Ref.~\cite{Brambilla:2020qwo}.

The resulting nuclear suppression $R_{AA}$ of each state is computed using
\be
R^{i}_{AA}(c,p_T,\phi) = \frac{\left(F \cdot S(c,p_T,\phi) \cdot \vec{\sigma}_{\text{direct}}\right)^{i}}{\vec{\sigma}_{\text{exp}}^{i}} \, ,
\label{eq:feeddown}
\ee
where $S(c,p_T,\phi)$ is a diagonal matrix which collects the survival probabilities extracted from the \qtraj evolution; $c$ labels the centrality class, $p_T$ the transverse momentum, and $\phi$ the azimuthal angle.
The experimental cross sections used are $\vec{\sigma}_{\text{exp}}=\{57.6$, 19, 3.72, 13.69, 16.1, 6.8, 3.27, 12.0, $14.15\}$ nb.
These values are computed from experimental measurements presented in Refs.~\cite{Sirunyan:2018nsz,Aaij:2014caa} as explained in Sec.~6.4 of Ref.~\cite{Brambilla:2020qwo}.

\begin{figure*}[t]
	\begin{center}
		\includegraphics[width=0.42\linewidth]{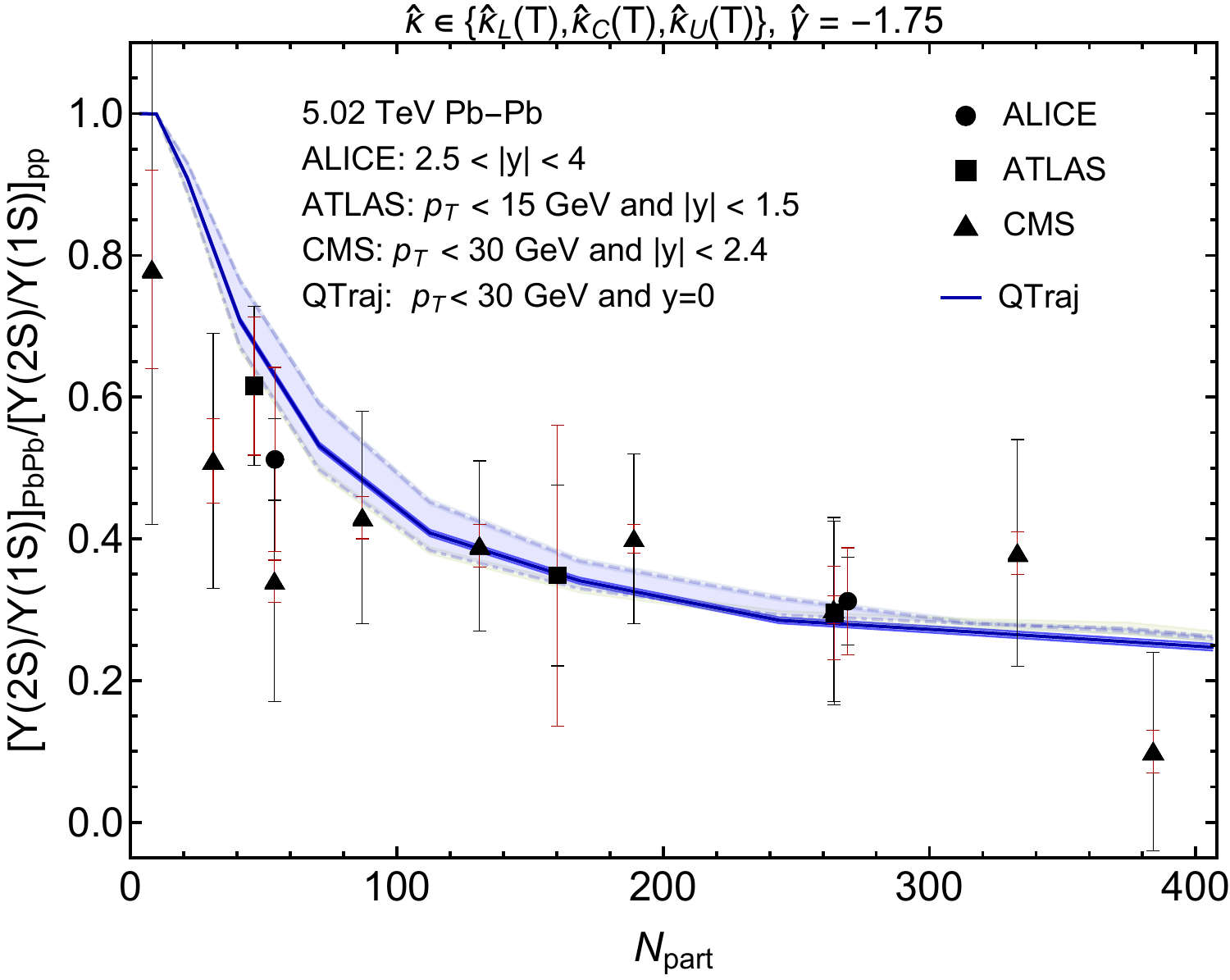} \hspace{1cm}
		\includegraphics[width=0.42\linewidth]{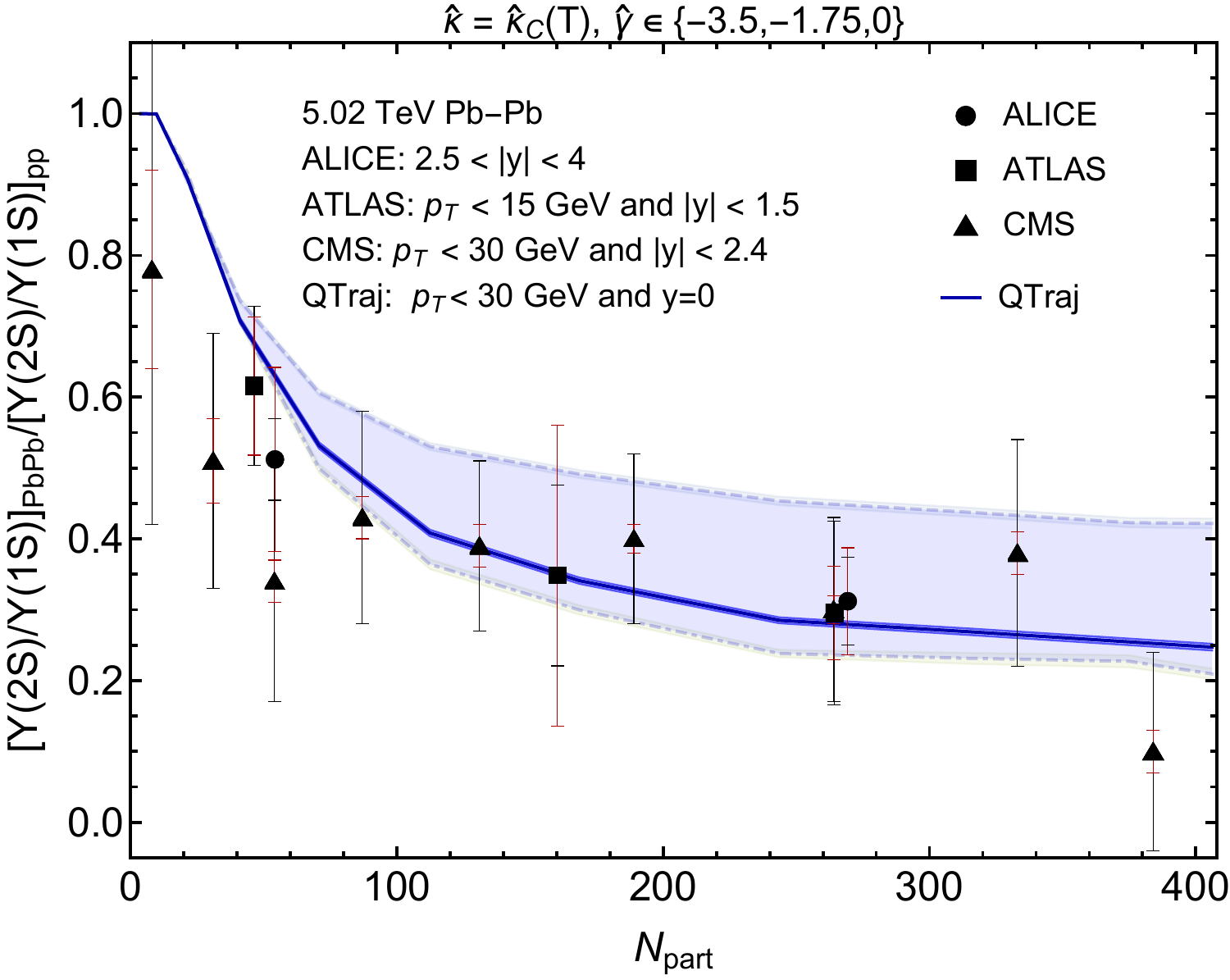}
	\end{center}
	\caption{(Color online)
		The double ratio of the nuclear modification factor $R_{AA}[\Upsilon(2S)]$ to $R_{AA}[\Upsilon(1S)]$ as a function of $N_{\text{part}}$ compared to experimental measurements of the ALICE~\cite{Acharya:2020kls}, ATLAS~\cite{ATLAS5TeV}, and CMS~\cite{CMS:2017ycw} collaborations.
		The bands in the theoretical curves indicate variation of $\hat{\kappa}(T)$ and $\hat{\gamma}$ as in Fig.~\ref{fig:raa_vs_npart}.
		The black and red bars in the experimental data represent statistical and systematic uncertainties, respectively.
	}
	\label{fig:2s_double_ratio_vs_npart}
\end{figure*}

\begin{figure*}[t]
	\begin{center}
		\includegraphics[width=0.42\linewidth]{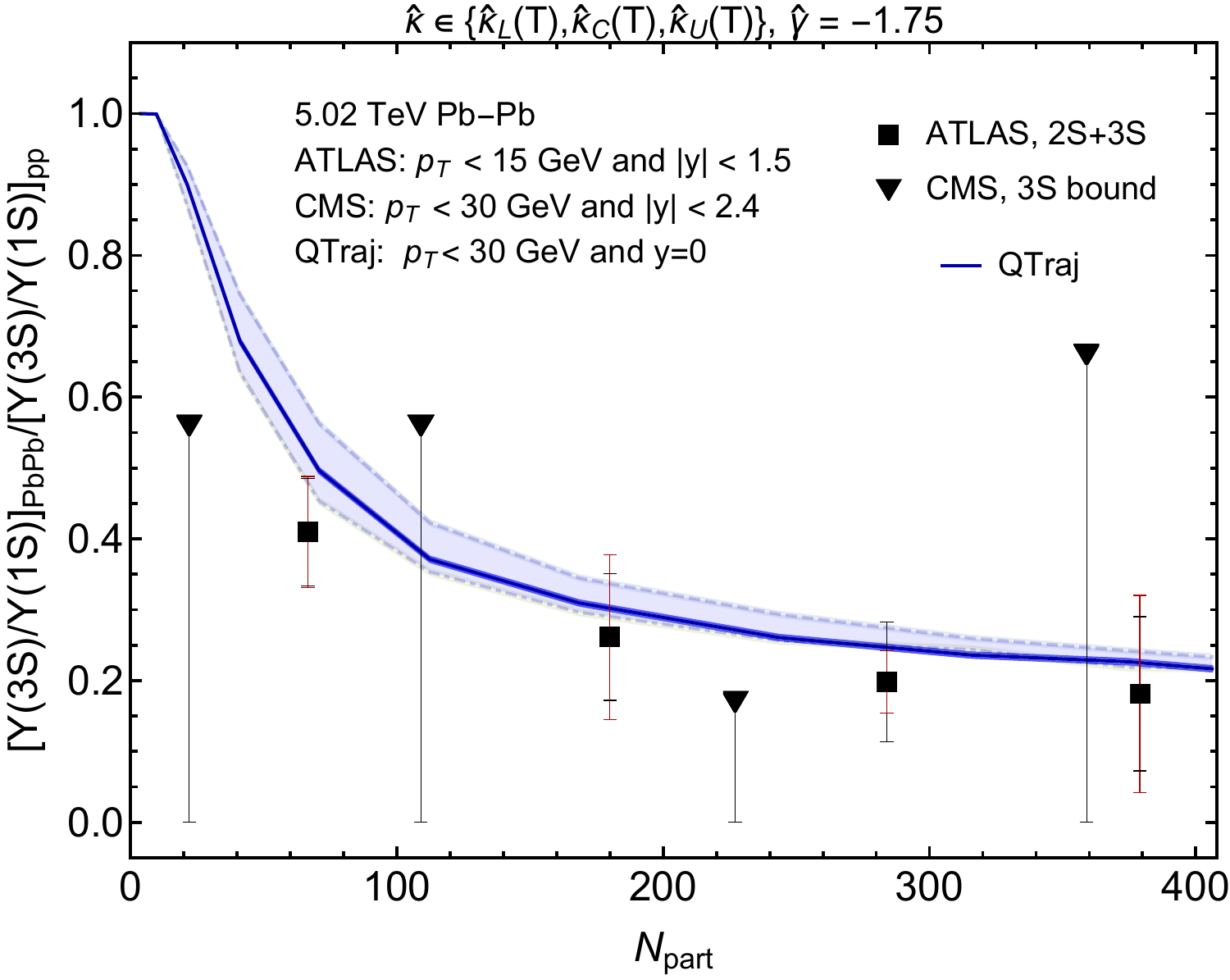}  \hspace{1cm}
		\includegraphics[width=0.42\linewidth]{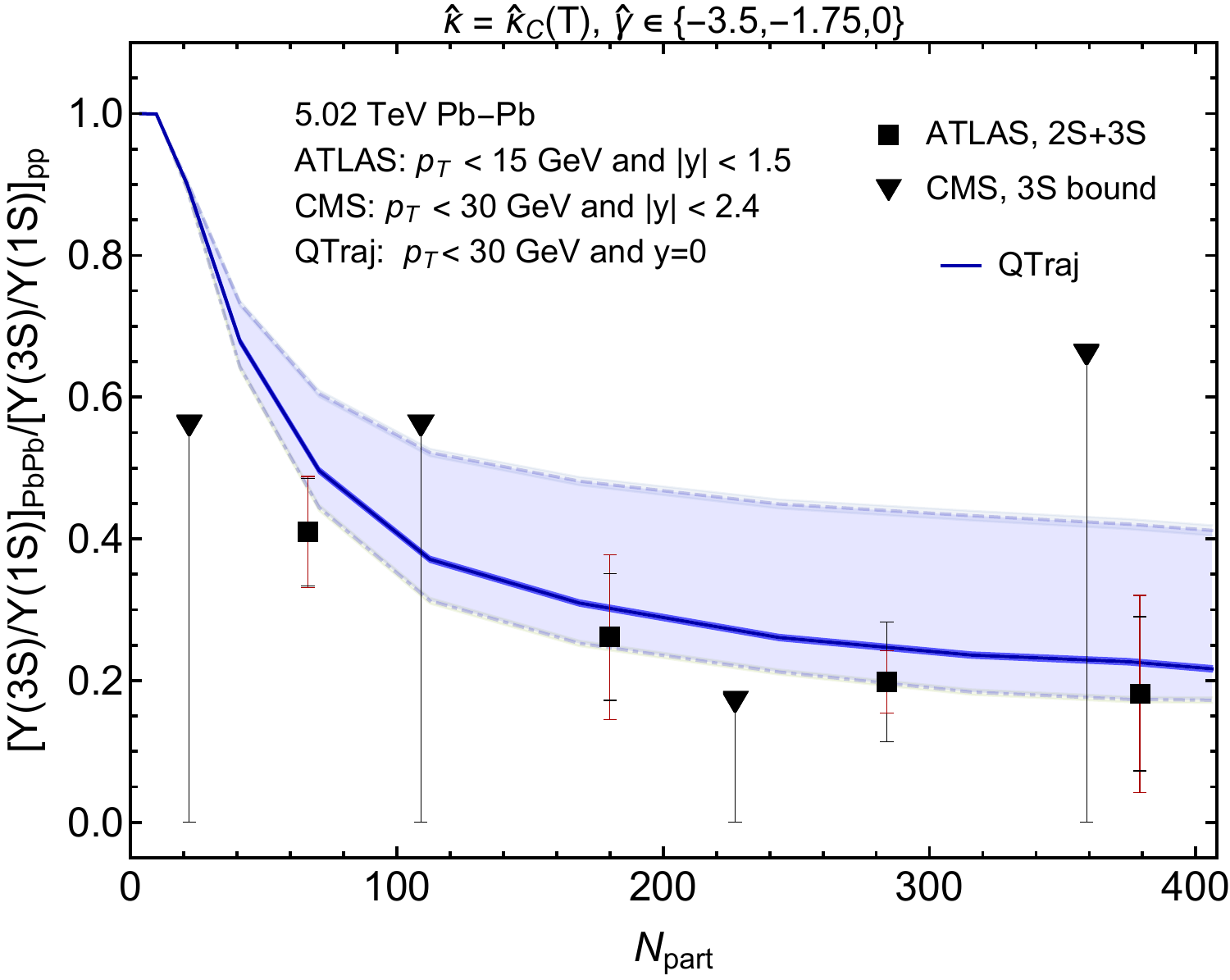}
	\end{center}
	\caption{(Color online)
		The double ratio of the nuclear modification factor $R_{AA}[\Upsilon(3S)]$ to $R_{AA}[\Upsilon(1S)]$ as a function of $N_{\text{part}}$ compared to experimental measurements of the ATLAS~\cite{ATLAS5TeV} and CMS~\cite{CMS:2017ycw} collaborations.
		The bands and bars represent uncertainties as in Fig.~\ref{fig:2s_double_ratio_vs_npart}; we note that the CMS measurements give only an upper bound at 95\% confidence level.
	}
	\label{fig:3s_double_ratio_vs_npart}
\end{figure*}

\section{Results} \label{sec:results}

In this section, we present our final results for the nuclear modification factor $R_{AA}$ and the elliptic flow $v_{2}$ of the $\Upsilon(1S)$, $\Upsilon(2S)$, and $\Upsilon(3S)$.
The theoretical uncertainties, which are indicated as shaded bands, come from varying the values of the parameters $\hat{\kappa}(T)$ and $\hat{\gamma}$ as detailed in Sec.~\ref{subsec:simulation_details}, while statistical errors are indicated by narrow bands around the individual lines which are, in many cases, smaller than the respective line widths.
In App.~\ref{app:jump_nojump}, for a subset of observables, we present comparisons between \qtraj simulations run with the full evolution including jumps as detailed in Sec.~\ref{subsec:qtraj} and results obtained by evolving the wave function using only the effective Hamiltonian $H_\text{eff} $ without applying the jump operators. 
The full \qtraj results presented in this section are obtained from approximately 50-100 quantum trajectories per 7\,-\,9\,$\times 10^5$ physical trajectories for each combination of $\hat\kappa(T)$ and $\hat\gamma$.
The $H_\text{eff} $ results were obtained by sampling approximately $10^{6}$ physical trajectories for each combination of $\hat\kappa(T)$ and $\hat\gamma$.
We compare our results with experimental data collected by the ALICE~\cite{Acharya:2020kls,Acharya:2019hlv}, ATLAS~\cite{ATLAS5TeV}, and CMS~\cite{Sirunyan:2018nsz,CMS:2017ycw,CMS:2020efs} collaborations.

\begin{figure*}[t]
	\begin{center}
		\includegraphics[width=0.42\linewidth]{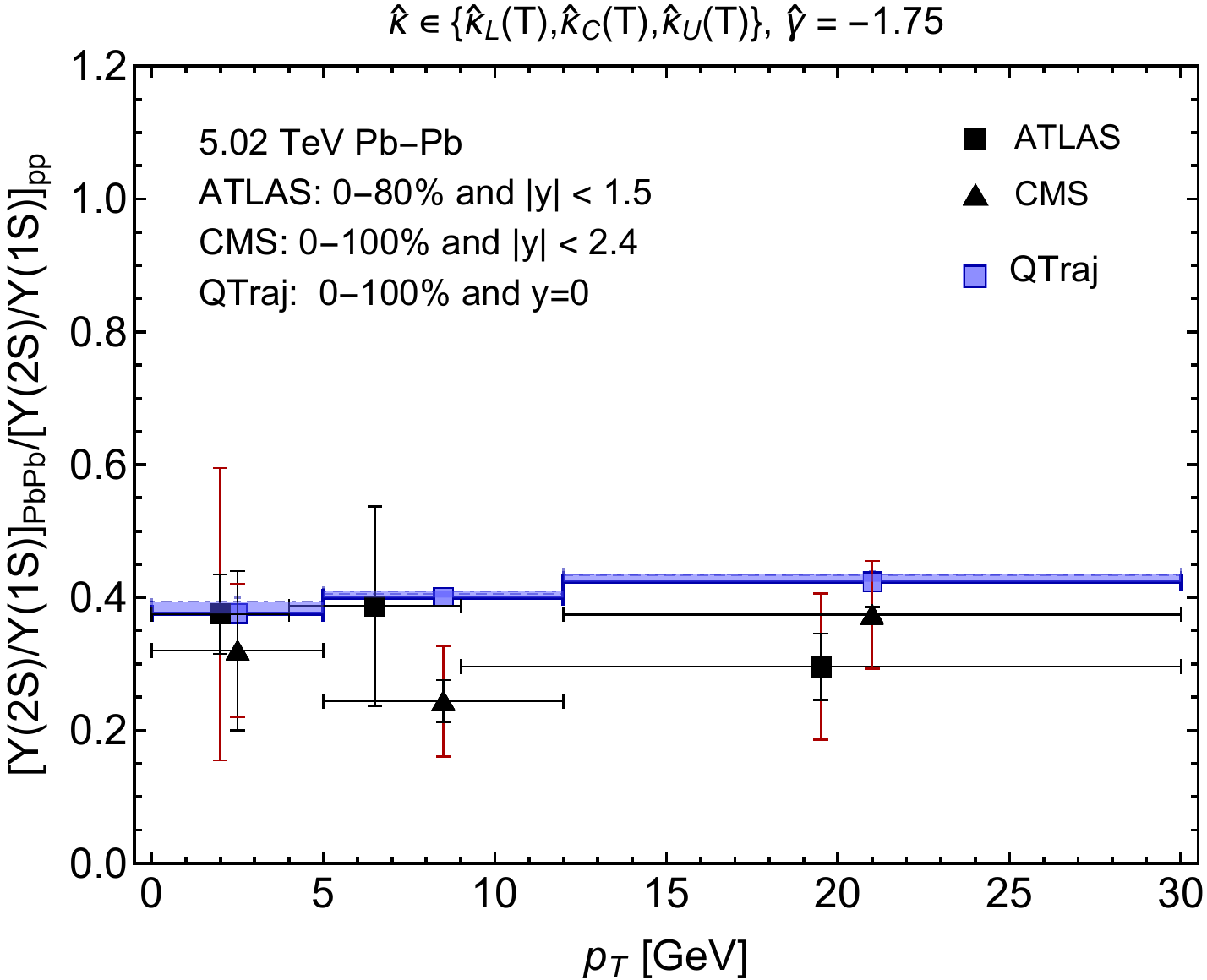}  \hspace{1cm}
		\includegraphics[width=0.42\linewidth]{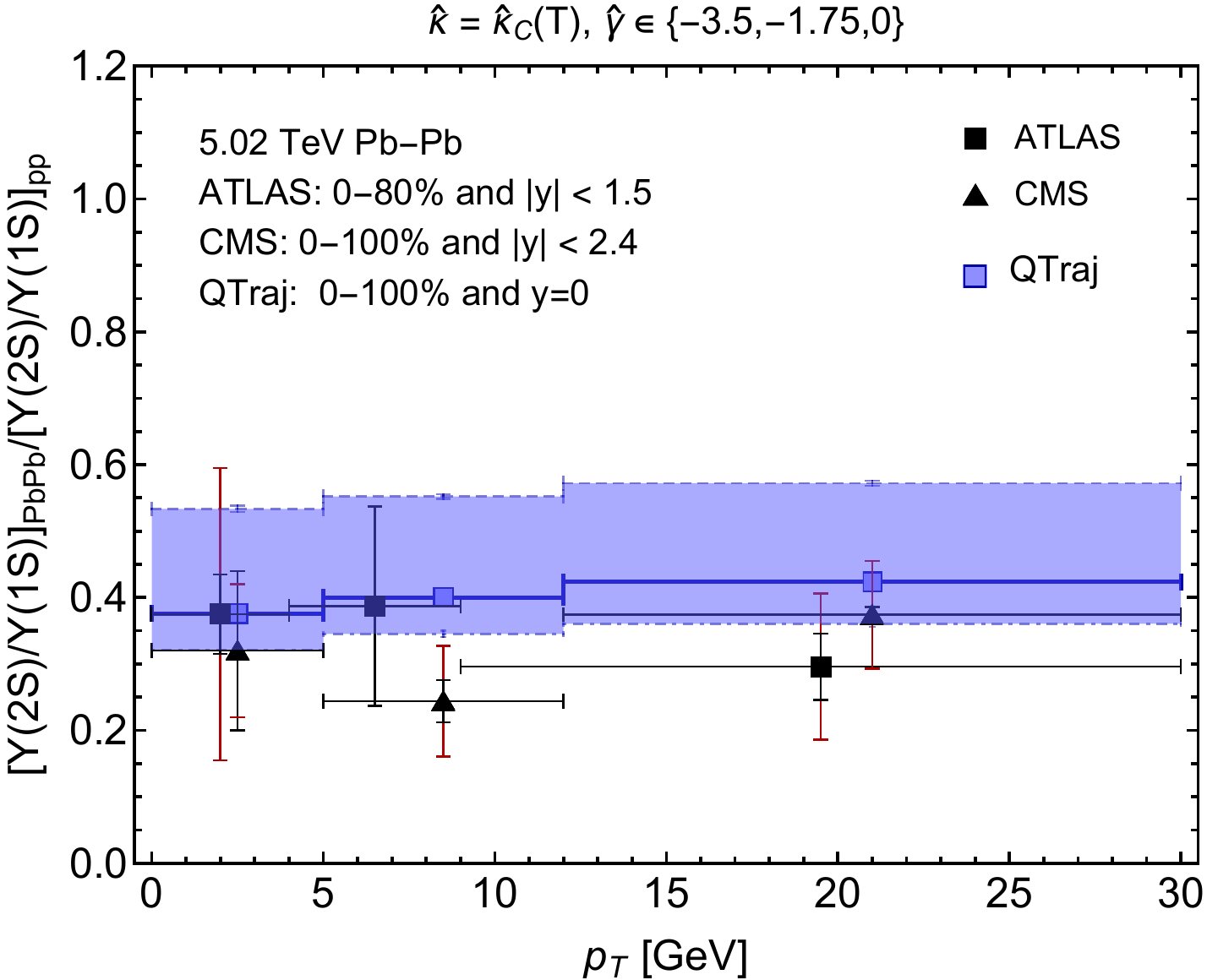}
	\end{center}
	\caption{(Color online)
		The double ratio of the nuclear modification factor $R_{AA}[\Upsilon(2S)]$ to $R_{AA}[\Upsilon(1S)]$ as a function of $p_{T}$ compared to experimental measurements of the ATLAS~\cite{ATLAS5TeV}, and CMS~\cite{CMS:2017ycw} collaborations.
		The bands and bars represent uncertainties as in Fig.~\ref{fig:2s_double_ratio_vs_npart}.
	}
	\label{fig:2s_double_ratio_vs_pt}
\end{figure*}

\begin{figure*}[t]
	\begin{center}
		\includegraphics[width=0.44\linewidth]{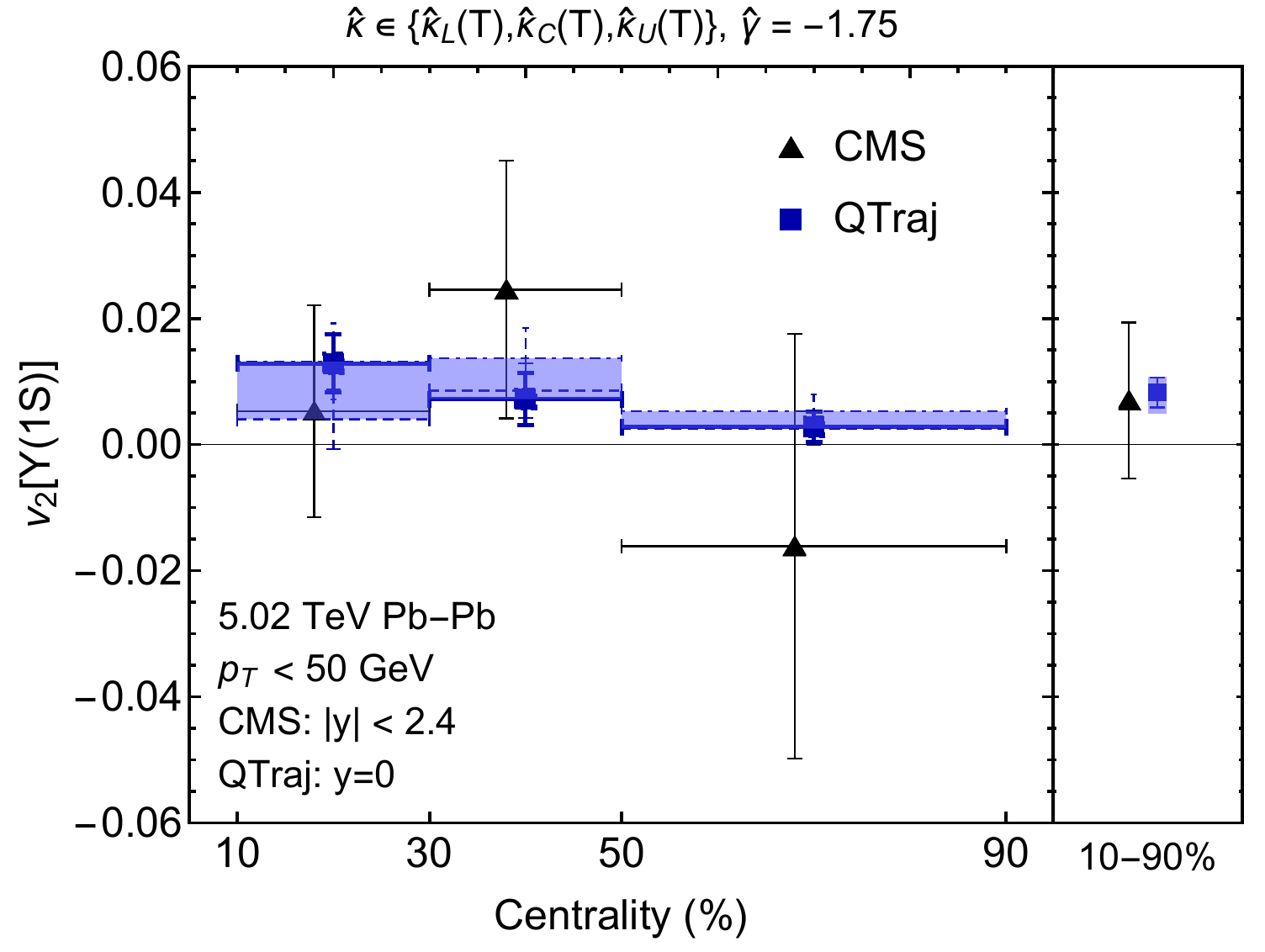} \hspace{5mm}
		\includegraphics[width=0.44\linewidth]{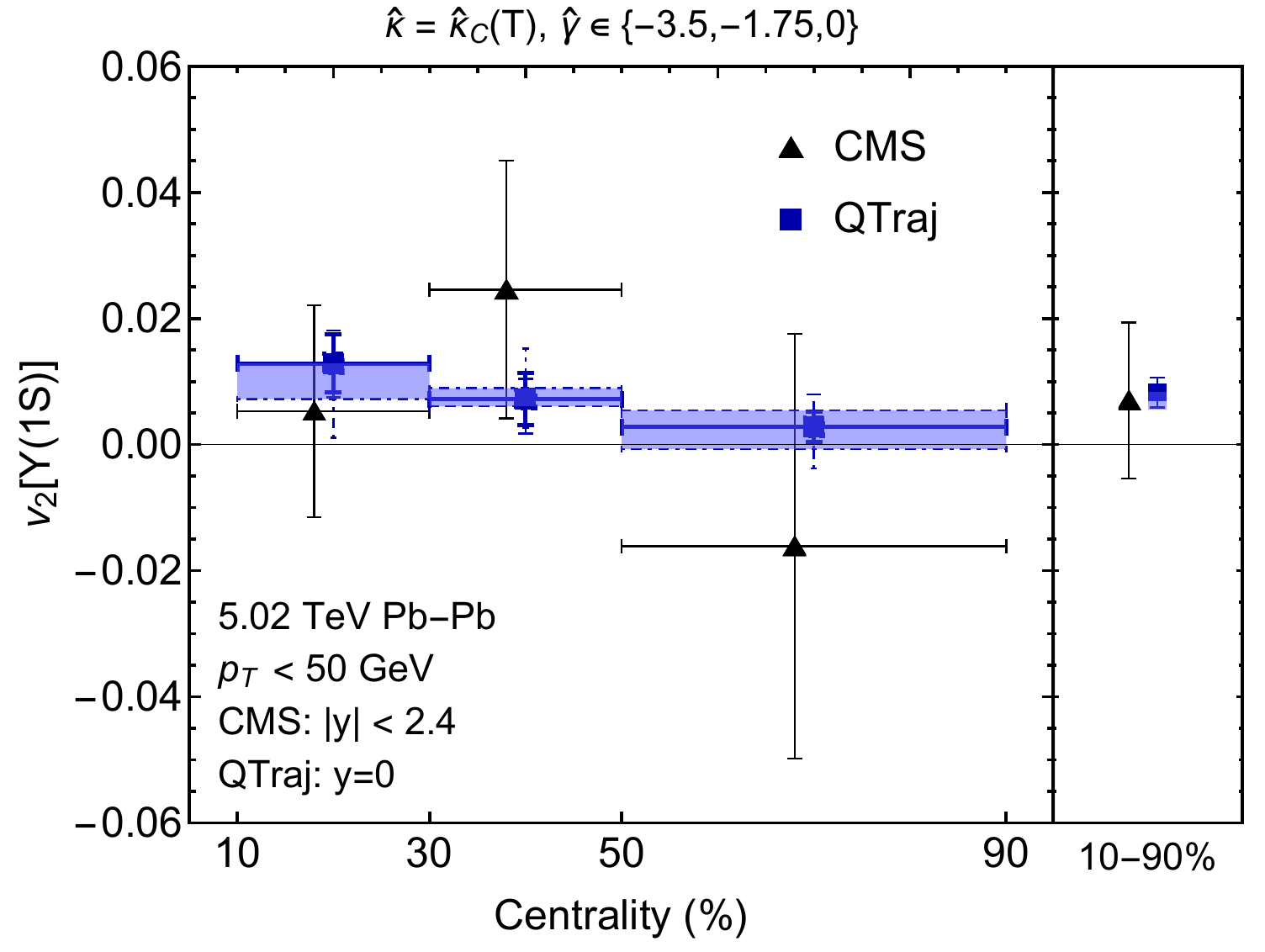}\;\;\;\;
	\end{center}
	\caption{(Color online)
		The elliptic flow $v_{2}$ of the $\Upsilon(1S)$ as a function of centrality compared to experimental measurements of the CMS~\cite{CMS:2020efs} collaboration.
		The bands represent uncertainties as in Fig.~\ref{fig:raa_vs_npart}.
	}
	\label{fig:v2_1S_vs_centrality}
\end{figure*}

\begin{figure*}[t]
	\begin{center}
		\includegraphics[width=0.42\linewidth]{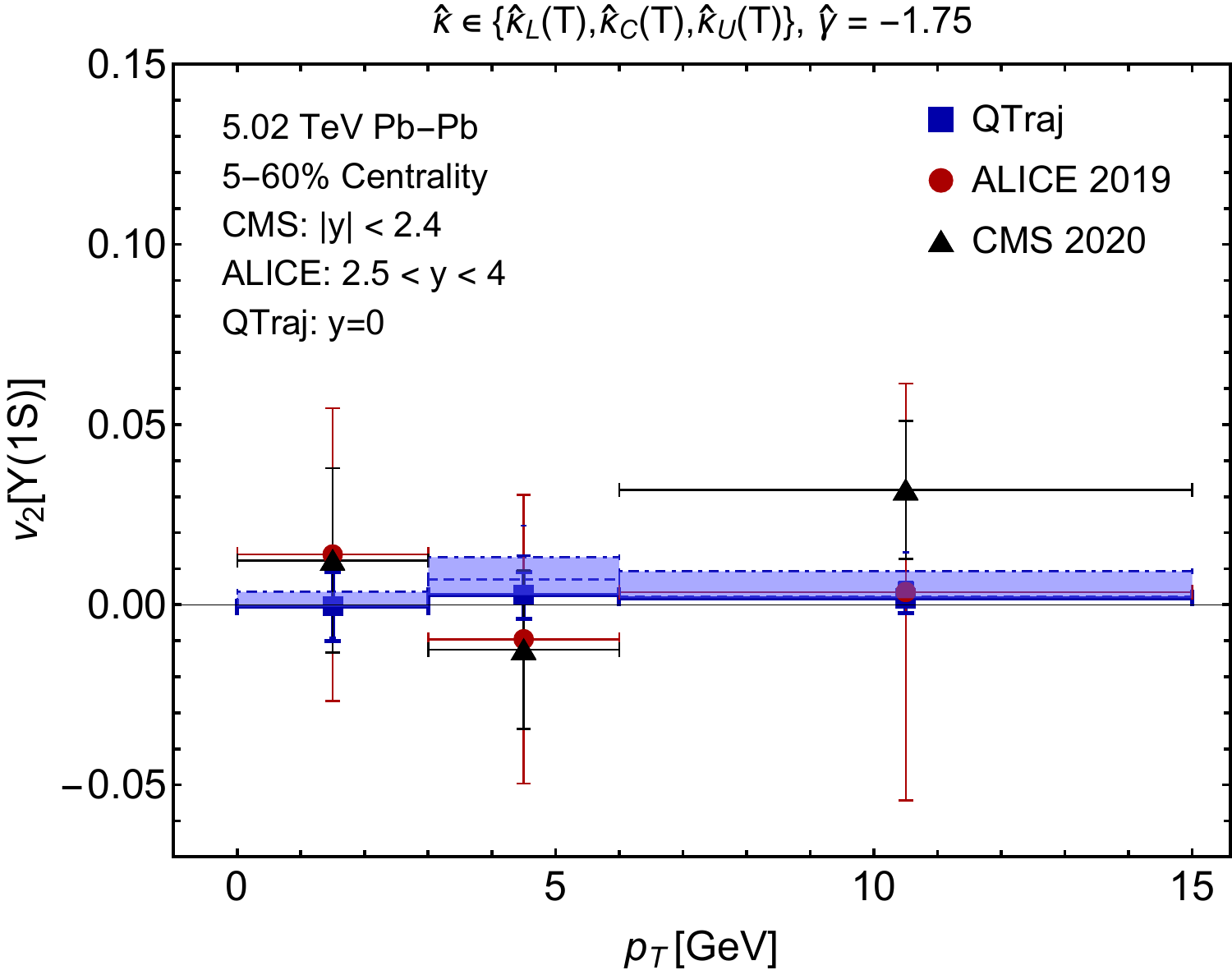} \hspace{1cm}
		\includegraphics[width=0.42\linewidth]{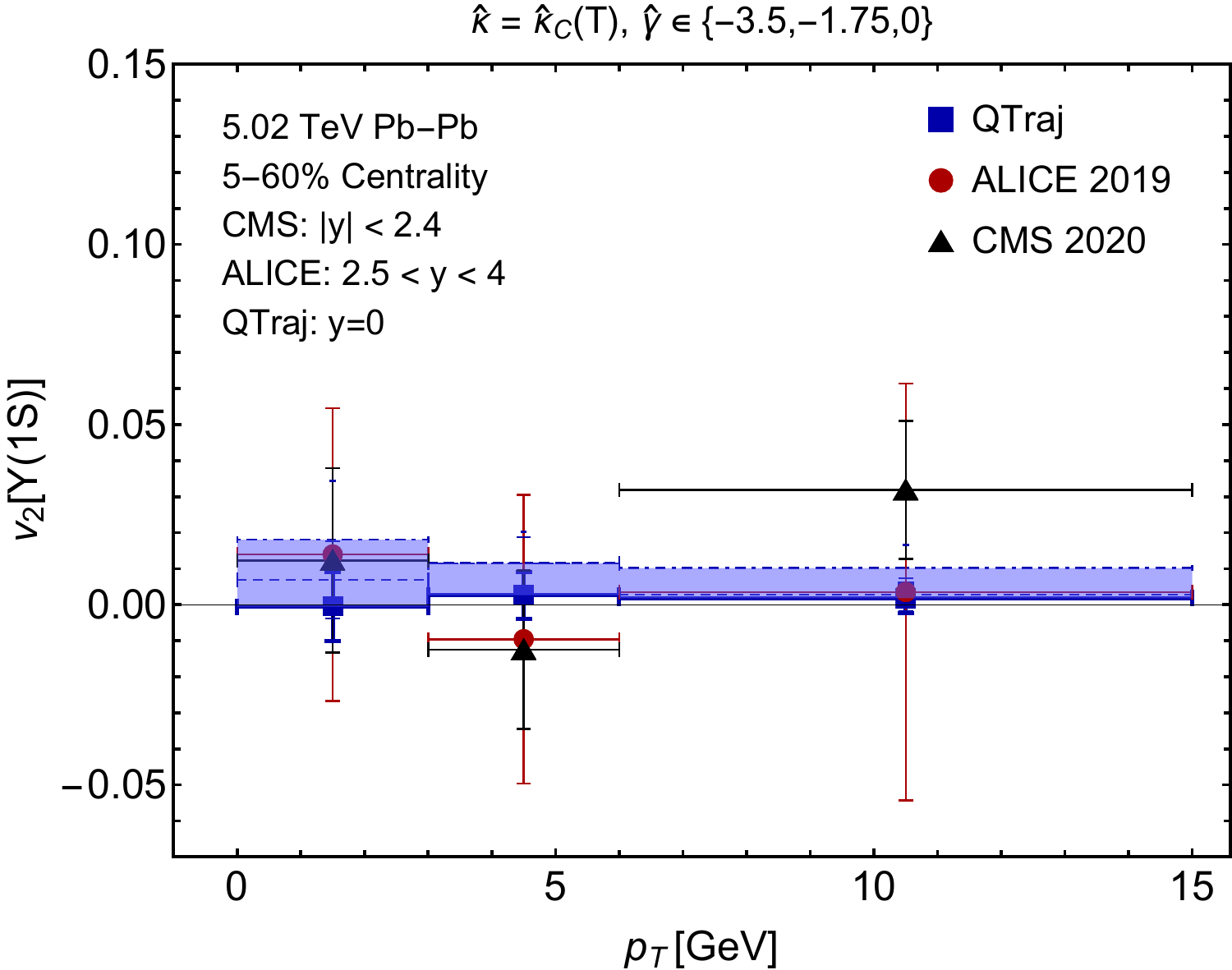}
	\end{center}
	\caption{(Color online)
		The elliptic flow $v_{2}$ of the $\Upsilon(1S)$ as a function of $p_{T}$ compared to experimental measurements of the ALICE~\cite{Acharya:2019hlv} and CMS~\cite{CMS:2020efs} collaborations.
		The bands represent uncertainties as in Fig.~\ref{fig:2s_double_ratio_vs_npart}.
		Note the much larger range of the ordinate compared to Fig.~\ref{fig:v2_1S_vs_centrality}.
	}
	\label{fig:v2_1S_vs_pt}
\end{figure*}

\begin{figure*}[t]
	\begin{center}
		\includegraphics[width=0.47\linewidth]{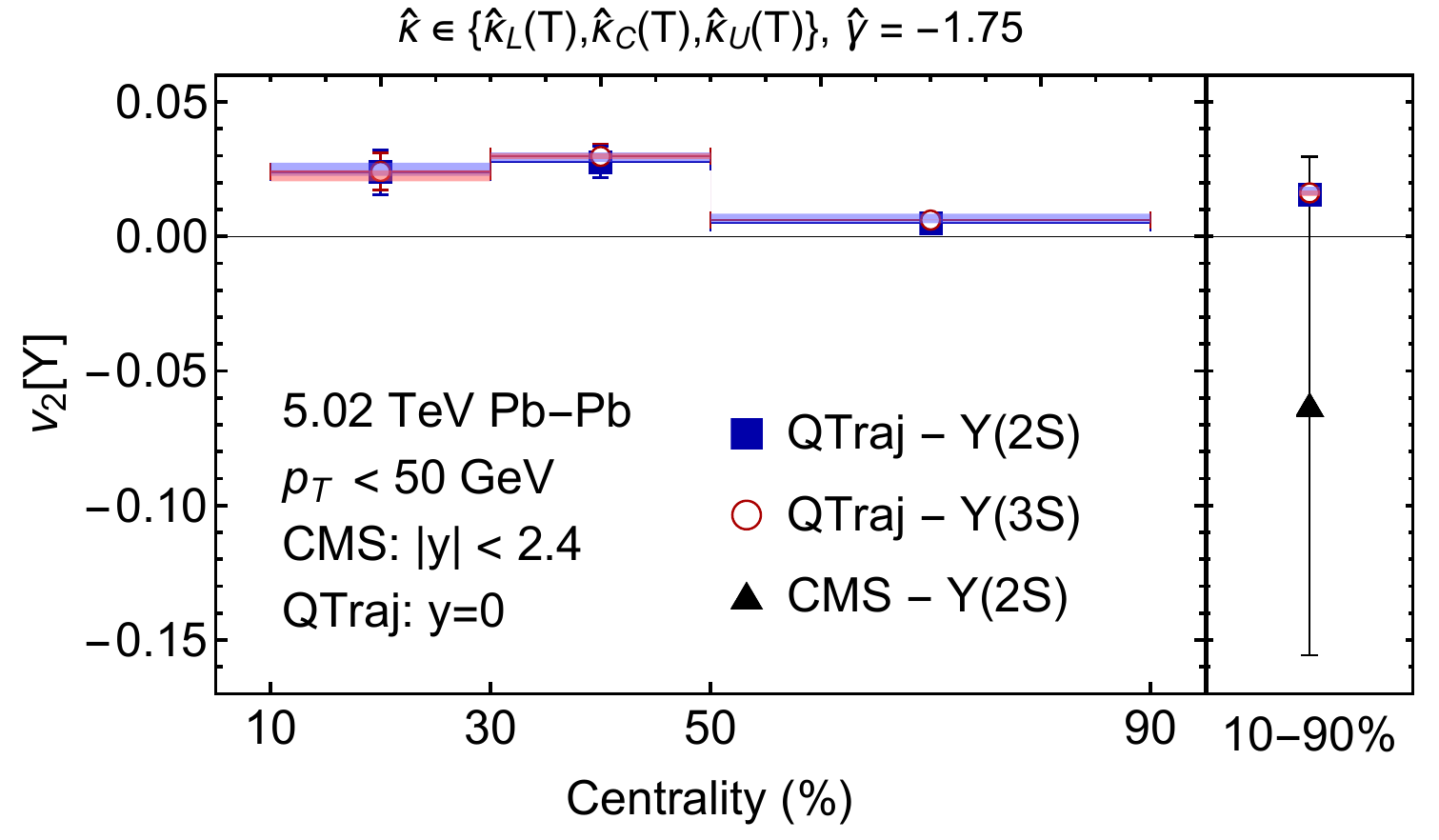} \hspace{5mm}
		\includegraphics[width=0.47\linewidth]{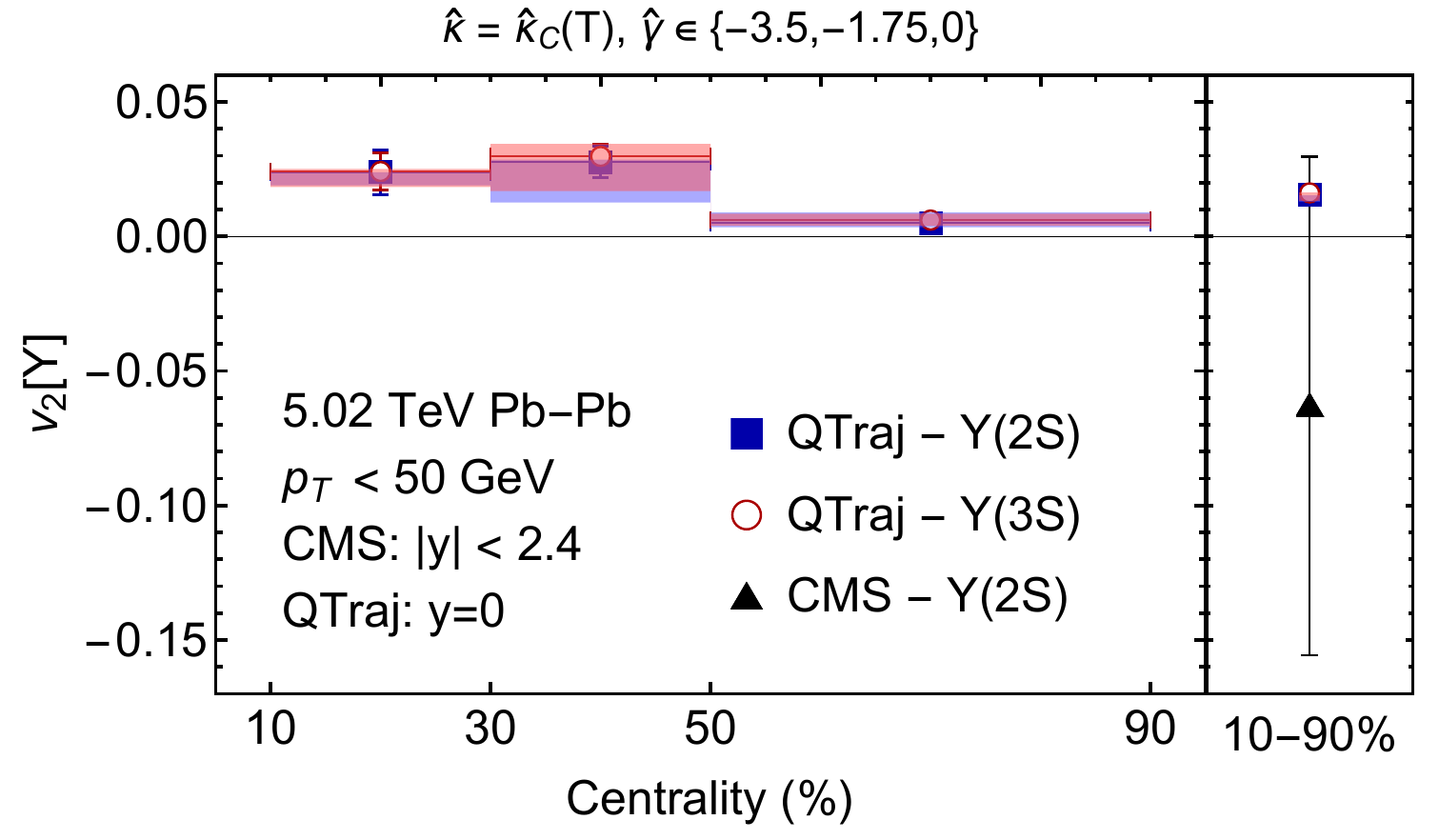}
	\end{center}
	\caption{(Color online)
		The elliptic flow $v_{2}$ of the $\Upsilon(2S)$ and $\Upsilon(3S)$ as a function of centrality compared to experimental measurements of the CMS~\cite{CMS:2020efs} collaboration.
		The bands represent uncertainties as in Fig.~\ref{fig:raa_vs_npart}.
	}
	\label{fig:v2_2S_3S_vs_centrality}
\end{figure*}

\subsection{Nuclear modification factor $R_{AA}$}

In Fig.~\ref{fig:raa_vs_npart}, we plot the results of our \qtraj simulations for the nuclear modification factor $R_{AA}$ of the $\Upsilon(1S)$, $\Upsilon(2S)$, and $\Upsilon(3S)$ as a function of the number of participating nucleons $N_{\text{part}}$.  
In the left panel of Fig.~\ref{fig:raa_vs_npart}, the shaded bands indicate the variation in our \qtraj results for $R_{AA}$ when varying $\hat\kappa$ while holding $\hat\gamma$ fixed at its central value; the dashed lines correspond to the lower bound $\hat{\kappa}(T) = \hat{\kappa}_L(T)$; and the dot-dashed lines correspond to the upper bound $\hat{\kappa}(T) = \hat{\kappa}_U(T)$.
In the right panel of Fig.~\ref{fig:raa_vs_npart}, the shaded bands indicate the variation in our \qtraj results for $R_{AA}$ when varying $\hat\gamma$ while holding $\hat\kappa$ fixed at its central value; the dashed lines correspond to the lower bound $\hat{\gamma} = -3.5$; and the dot-dashed lines correspond to the upper bound $\hat{\gamma} = 0$.
As can be seen from this figure, the central values of these two parameters provide a good description of the $N_\text{part}$ dependence of $R_{AA}$ for all three states considered.  
Comparing the left and right panels of Fig.~\ref{fig:raa_vs_npart}, one sees that the uncertainty associated with the variation of $\hat\gamma$ (right panel)
is larger than the one  associated with the variation of $\hat\kappa$ (left panel).

In Fig.~\ref{fig:raa_vs_pt}, we present our results for $R_{AA}[1S]$, $R_{AA}[2S]$, and $R_{AA}[3S]$ as a function of transverse momentum $p_{T}$. 
The bands, line styles, and panels represent the same variation as in Fig.~\ref{fig:raa_vs_npart}.
We observe that, within uncertainties, our results are in agreement with the experimental data. 
In fact, the dependence of $R_{AA}$ on $p_T$ is very mild. 
This behavior is seen both in our results and in experimental measurements. 
Our results for $\Upsilon(1S)$ show a greater sensitivity to variation of $\hat{\gamma}$ than to $\hat{\kappa}$; however, the opposite is true for the excited states. 

In Figs.~\ref{fig:2s_double_ratio_vs_npart} and \ref{fig:3s_double_ratio_vs_npart}, we present our results for the double ratio of $R_{AA}[2S]$ and $R_{AA}[3S]$, respectively, to $R_{AA}[1S]$ as a function of $N_{\text{part}}$.  
As in Fig.~\ref{fig:raa_vs_npart}, the left and right panels correspond to the variation over $\hat\kappa(T)$ and $\hat\gamma$, and the line styles for the bounds are the same.
We note that the data from the CMS collaboration in Fig.~\ref{fig:3s_double_ratio_vs_npart} give only an upper bound on $R_{AA}[3S]$.
We observe good agreement between our \qtraj results and the experimentally measured values of the double ratios across the entire range of $N_{\text{part}}$. 
Our results show a much larger dependency on $\hat{\gamma}$ than on $\hat{\kappa}$. 
This suggests that this measurement can potentially constrain the value of $\hat{\gamma}$, for which there are much less lattice QCD data than for $\hat{\kappa}$. Unfortunately, at the moment, the experimental uncertainties are of the order of the effect of the $\hat{\gamma}$ variation.

In Fig.~\ref{fig:2s_double_ratio_vs_pt}, we plot the double ratio of $R_{AA}[2S]$ to $R_{AA}[1S]$ as a function of $p_{T}$. 
The notation and parameter variation are the same as in the previous plots.
What we observe in this figure confirms what we saw in previous plots. 
The dependence of this double ratio with $p_T$ is very mild. 
And similarly to what we observed in the double ratio versus the number of participants, varying $\hat{\kappa}$ has almost no influence while varying $\hat{\gamma}$ is significant. 
Regarding the comparison with experimental data, we see a reasonable agreement within reported uncertainties with some tension with the data seen at large $p_T$.

\subsection{Elliptic flow $v_{2}$}

In Fig.~\ref{fig:v2_1S_vs_centrality}, we plot our results for the elliptic flow $v_{2}$ of the $\Upsilon(1S)$ as a function of centrality. 
Again, the notation is as in previous plots. 
Our results agree to within uncertainties with the experimental results of the CMS collaboration, though we note the large uncertainities of the experimental results. 
In this case, we see that the influence of $\hat{\kappa}$ and $\hat{\gamma}$ is similar. 
It is noteworthy that the more inclusive prediction (in the $10$ to $90\%$ percent centrality window) is very precise and close to the central value of the experimental results (more details below).

In Fig.~\ref{fig:v2_1S_vs_pt}, we plot $v_{2}[\Upsilon(1S)]$ as a function of $p_{T}$.
We observe agreement to within uncertainties with the experimental results of the ALICE and CMS collaborations. 
In this case, the sensitivity of our results to $\hat{\kappa}$ and $\hat{\gamma}$ is similar, except for the lower momentum region, in which the sensitivity to $\hat{\gamma}$ is larger. 

In Fig.~\ref{fig:v2_2S_3S_vs_centrality}, we plot our results for the elliptic flow of the $\Upsilon(2S)$ and $\Upsilon(3S)$ as a function of centrality.
Our results agree to within uncertainties with the experimental data point from the CMS collaboration, although the experimental uncertainties are at least an order of magnitude larger than our theoretical uncertainty. It is interesting to see that our model predicts very similar $v_2$ for both $\Upsilon(2S)$ and $\Upsilon(3S)$; $v_2$ for the excited states appears to be somewhat larger than for the ground state. 

In the case of $v_2$ of the $\Upsilon(1S)$, we predict that it has a maximum on the order of 1.5\% as a function of both centrality and transverse momentum.  
Our prediction for the 10-90\% centrality- and $p_T$-integrated $\Upsilon$ elliptic flow is $v_2[\Upsilon(1S)]=0.008 \pm 0.003 \pm 0.002$, $v_2[\Upsilon(2S)]=0.016 \pm 0.003 \pm 0.002$, and $v_2[\Upsilon(3S)]=0.015 \pm 0.002 \pm 0.001$, where the first uncertainty corresponds to both $\hat\kappa$ and $\hat\gamma$ variation and the second uncertainty corresponds to the statistical uncertainty due to the average over physical and quantum trajectories.  
We find that the 2S and 3S states have similar integrated elliptic flow, which is roughly a factor of two larger than the 1S state, $v_2[\Upsilon(2S\ \text{or}\ 3S)]/v_2[\Upsilon(1S)] \simeq 2$ for 10-90\%.  
When considering the 2S to 1S $v_2$-ratio in different centrality bins, we find that, taking into account the variation over both $\hat\kappa$ and $\hat\gamma$ results in $2 \lesssim v_2[\Upsilon(2S)]/v_2[\Upsilon(1S)]  \lesssim 4$, with the maximum in this ratio occurring in the 30-50\% centrality bin.

\section{Conclusions and outlook}
\label{sec:conclusions}

In this paper, we presented a comprehensive set of predictions for the suppression and elliptic flow of $\Upsilon(1S)$, $\Upsilon(2S)$, and $\Upsilon(3S)$ in 5 TeV Pb-Pb collisions and compared our predictions to experimental data from the ALICE, ATLAS, and CMS experiments.  
To make our predictions, we numerically solved the 3D non-abelian Lindblad equation for the quarkonium reduced density matrix that emerges when OQS methods are applied within the pNRQCD effective field theory for a strongly-coupled QGP.  
The numerical solution was realized by mapping the solution of the Lindblad equation to a 1D Schr\"odinger equation with a non-Hermitian Hamiltonian that is subject to stochastic quantum jumps.  
Using the resulting quantum trajectories algorithm, we were able to simulate the full 3D evolution of the wave-function, including the possibility of internal transitions between different color and angular momentum states.

To describe the interaction with the hot and three-dimensionally expanding QGP, we made use of a realistic dissipative hydrodynamics simulation called anisotropic hydrodynamics.  
The initial conditions and transport coefficients used in the 3+1D aHydro code were tuned to reproduce soft observables such as identified pion, proton, and kaon $p_T$-spectra, multiplicities, and elliptic flow.  
To compute $R_{AA}$, we produced a large ensemble of physical quarkonium trajectories by Monte-Carlo sampling both the initial production points and transverse momentum vectors.  
We then computed the survival probability along each of these physical trajectories by averaging over ensembles of stochastically generated quantum trajectories.  
Based on the Monte-Carlo sampling of physical trajectories, we could compute both the $N_{\rm part}$- and $p_T$-dependence of $R_{AA}$ and the elliptic flow of the states.  
This extends our prior work where, due to the high computational demand of solving the Lindblad equation, we used a trajectory-averaged temperature evolution in each centrality bin \cite{Brambilla:2020qwo}.
Our final predictions also include the effect of late-time feed down of bottomonium states, the calculation of which is based on known experimental measurements of bottomonium production cross-sections and branching ratios in pp collisions.  
We find that the primary effect of computing the survival probability on a trajectory-by-trajectory basis is to increase both $\Upsilon(2S)$ and $\Upsilon(3S)$ $R_{AA}$, which helps to bring our predictions for both $R_{AA}$ of these states, and the corresponding double ratios, into better agreement with available experimental data than the trajectory-averaged results presented in \cite{Brambilla:2020qwo}.  Associated with this paper, the \qtraj code used to generate the results will be released under a public GPL license.  We present the details of the code, along with examples, and benchmarks in a separate work with a more computational focus \cite{compforth}.

Due to the stochastic quantum trajectories algorithm and Monte-Carlo sampling of the physical trajectories, the results of our simulation had an associated statistical uncertainty.  
For each parameter set considered, the statistical uncertainty computed was reported in each figure based on an ensemble size of approximately $10^5{-}10^6$ physical trajectories.  
With these large ensemble sizes, the statistical uncertainty in the determination of $R_{AA}$ was on the order of the line width in the plots, while there remained somewhat larger statistical uncertainties in our predictions for $v_2$.  
We estimated our theoretical uncertainties by varying the relevant transport coefficients $\hat\kappa$ and $\hat\gamma$ in the range indicated by lattice measurements of these quantities.  
We found that $R_{AA}[\Upsilon(1S)]$, the 2S to 1S double ratio, and 3S to 1S double ratio had a larger variation with $\hat\gamma$ than with $\hat\kappa$, with the double-ratios rather strongly depending on $\hat\gamma$ but not $\hat\kappa$.  
This observation offers some hope that, with increased statistics for both 1S and 2S $R_{AA}$, one can constrain $\hat\kappa$ and $\hat\gamma$ based on experimental data.

In the case of the elliptic flow, we found similar variation in our predictions under variation of $\hat\kappa$ and $\hat\gamma$.  
We found reasonable agreement between our predictions for $v_2[\Upsilon(1S)]$ and available experimental data and made predictions for the elliptic flow of the $\Upsilon(2S)$ and $\Upsilon(3S)$, finding that the differential suppression of these states results in a larger elliptic flow, as can be expected from the fact that their survival probabilities are smaller (stronger medium interactions).  
When considering the centrality dependence of the ratio of the elliptic flow of the 2S and 1S states, our approach predicts $2 \lesssim v_2[\Upsilon(2S)]/v_2[\Upsilon(1S)]  \lesssim 4$, with the maximum occurring in the 30-50\% centrality bin.  
This prediction can hopefully soon be tested by experimentalists.

Turning to the future, one limitation of the framework used herein is that it relied on an assumed strict ordering of the binding energy and temperature, namely $T \gg E$.  
As a result, at low-temperatures, the framework used herein becomes potentially unreliable.  
For this reason, we used a lower temperature of $T_f = 250$ MeV for bottomonium interactions with the medium.  
In our previous work, it was shown that the variation of $R_{AA}$ when varying $T_f$ by 10\% was on the same order as the theoretical uncertainty associated with the variation of the fundamental transport coefficients $\hat\kappa$ and $\hat\gamma$.  
That said, it seems necessary to include sub-leading corrections in $E/T$ in order to gauge their impact on in-medium bottomonium dynamics~\cite{Akamatsu:2020ypb}.  
Another interesting prospect is that, at low temperatures, one could interface \qtraj output to codes based on a semi-classical approach in which one instead solves in-medium Boltzmann equations, see e.g. \cite{Yao:2018nmy,Yao:2020xzw,Yao:2020eqy}.

\acknowledgments{
	N.B., P.V. and A.V. acknowledge support by the DFG cluster of excellence ORIGINS funded by the Deutsche Forschungsgemeinschaft under Germany's Excellence Strategy - EXC-2094-390783311.
	This work has also received financial support from Xunta de Galicia (Centro singular de investigaci\'{o}n de Galicia accreditation 2019-2022), by European Union ERDF, by  the ``Mar\'{i}a  de Maeztu''  Units  of  Excellence program  MDM-2016-0692, the Spanish Research State Agency and from the European Research Council project ERC-2018-ADG-835105 YoctoLHC.
	J.H.W.'s research has been also funded by the DFG - Projektnummer 417533893/GRK2575 ``Rethinking Quantum Field Theory''.
	M.S. has been supported by the U.S. Department of Energy, Office of Science, Office of Nuclear Physics Award No.~DE-SC0013470.
	M.S. also thanks the Ohio Supercomputer Center for support under the auspices of Project No.~PGS0253.  
}

\appendix

\section{Table of results for integrated $R_{AA}$}
\label{app:table}

In Tab.~\ref{tab:comp}, we present \qtraj predictions for the integrated $R_{AA}$ of 1S, 2S, and 3S along with the corresponding results from the ALICE, ATLAS, and CMS experiments.  We note that, for the \qtraj results, the variation over the full $\hat\gamma$ range was the dominant source of systematic theoretical uncertainty in all cases listed.

\begin{table}[ht]
	\begin{center}
		\def\arraystretch{1.5}
		{\scriptsize
			\begin{tabular}{|c|c|c|c|}
				\hline
				{\bf ~Observable~} & {\bf ~Source/Cuts~}  &{\bf ~Experiment/\qtraj~}  \\
				\hline
				$R_{AA}[\Upsilon(1S)]$ & ALICE 0-90\%  \cite{Acharya:2018mni}& 0.37 $\pm$ 0.03 $\pm$ 0.02  \\
				& $p_T < 15$ GeV & $0.35 \pm 0.09 \pm 0.002$  \\
				\hline
				$R_{AA}[\Upsilon(1S)]$ & ATLAS 0-80\% \cite{ATLAS5TeV} & 0.32 $\pm$ 0.05 $\pm$ 0.02    \\
				& $p_T < 30$ GeV & $0.35 \pm 0.09 \pm 0.002$ \\
				\hline
				$R_{AA}[\Upsilon(1S)]$ & CMS 0-100\% \cite{Sirunyan:2018nsz}& 0.376 $\pm$ 0.035 $\pm$ 0.013   \\
				& $p_T < 30$ GeV & $0.36 \pm 0.09 \pm 0.002$ \\
				\hline
				$R_{AA}[\Upsilon(2S)]$ & ALICE 0-90\% \cite{Acharya:2018mni}& 0.10 $\pm$ 0.02 $\pm$ 0.04 \\
				& $p_T < 15$ GeV & $0.139 \pm 0.022 \pm 0.001$  \\
				\hline
				$R_{AA}[\Upsilon(2S)]$ & ATLAS 0-80\% \cite{ATLAS5TeV} & 0.11 $\pm$ 0.04 $\pm$ 0.04   \\
				& $p_T < 30$ GeV & $0.137 \pm 0.022 \pm 0.001$  \\
				\hline
				$R_{AA}[\Upsilon(2S)]$ & CMS 0-100\% \cite{Sirunyan:2018nsz}& 0.117 $\pm$ 0.019 $\pm$ 0.022   \\
				& $p_T < 30$ GeV & $0.148 \pm 0.022 \pm 0.001$ \\
				\hline
				$R_{AA}[\Upsilon(3S)]$ & CMS 0-100\% \cite{Sirunyan:2018nsz}& 0.022 $\pm$ 0.016 $\pm$ 0.038    \\
				& $p_T < 30$ GeV & $0.138 \pm 0.008 \pm 0.001$ \\
				\hline
			\end{tabular}
		}
	\end{center}
	\caption{Comparison of \qtraj predictions for integrated $R_{AA}[\Upsilon]$ with available experimental data.  In the right column, the top value is the experimental value, and the bottom value is the \qtraj prediction.  
	With the exception of $R_{AA}[\Upsilon(3S)]$, results agree within quoted uncertainties.
	In all cases, the first uncertainty quoted is the systematic uncertainty, and the second is the statistical uncertainty.
	}
	\label{tab:comp}
\end{table}

\begin{figure*}[t]
	\begin{center}
		\includegraphics[width=0.94\linewidth]{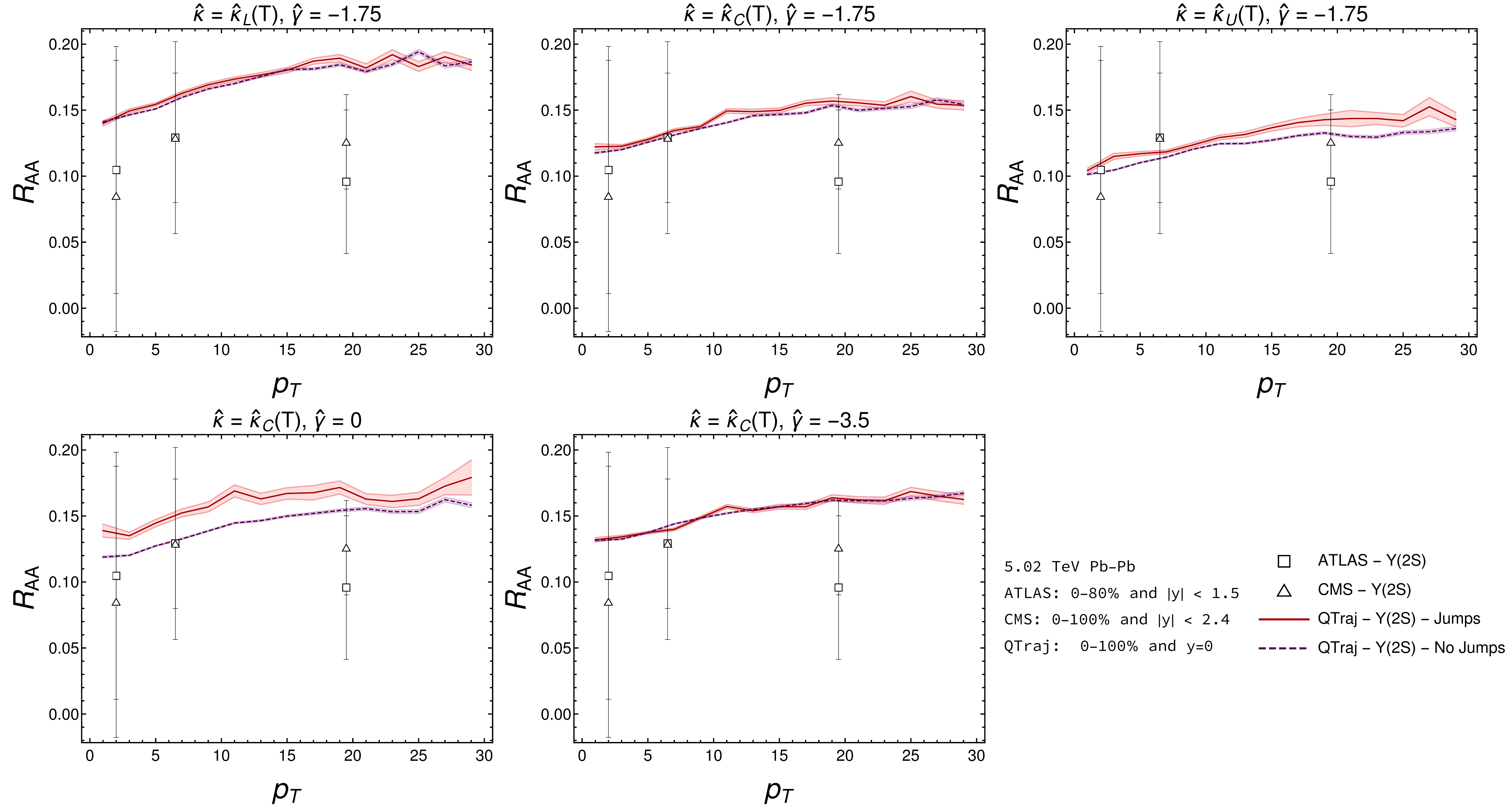}
	\end{center}
	\caption{(Color online)
		The nuclear modification factor $R_{AA}$ of the $\Upsilon(2S)$ as a function of $p_{T}$ together with experimental measurements of $R_{AA}[\Upsilon(2S)]$ from the ATLAS~\cite{ATLAS5TeV}, and CMS~\cite{Sirunyan:2018nsz} collaborations.
		We compare the results obtained using the full \qtraj algorithm (red, solid) with results obtained using evolution with $H_\text{eff}$ with no jumps (purple, dashed).
		The top row varies $\hat{\kappa}(T)$ at $\hat{\gamma}=-1.75$, and the bottom row varies $\hat{\gamma}$ at $\hat{\kappa}_{C}(T)$.
	}
	\label{fig:raa2S_vs_pt_jump_comp}
\end{figure*}

\begin{figure*}[t]
	\begin{center}
		\includegraphics[width=0.94\linewidth]{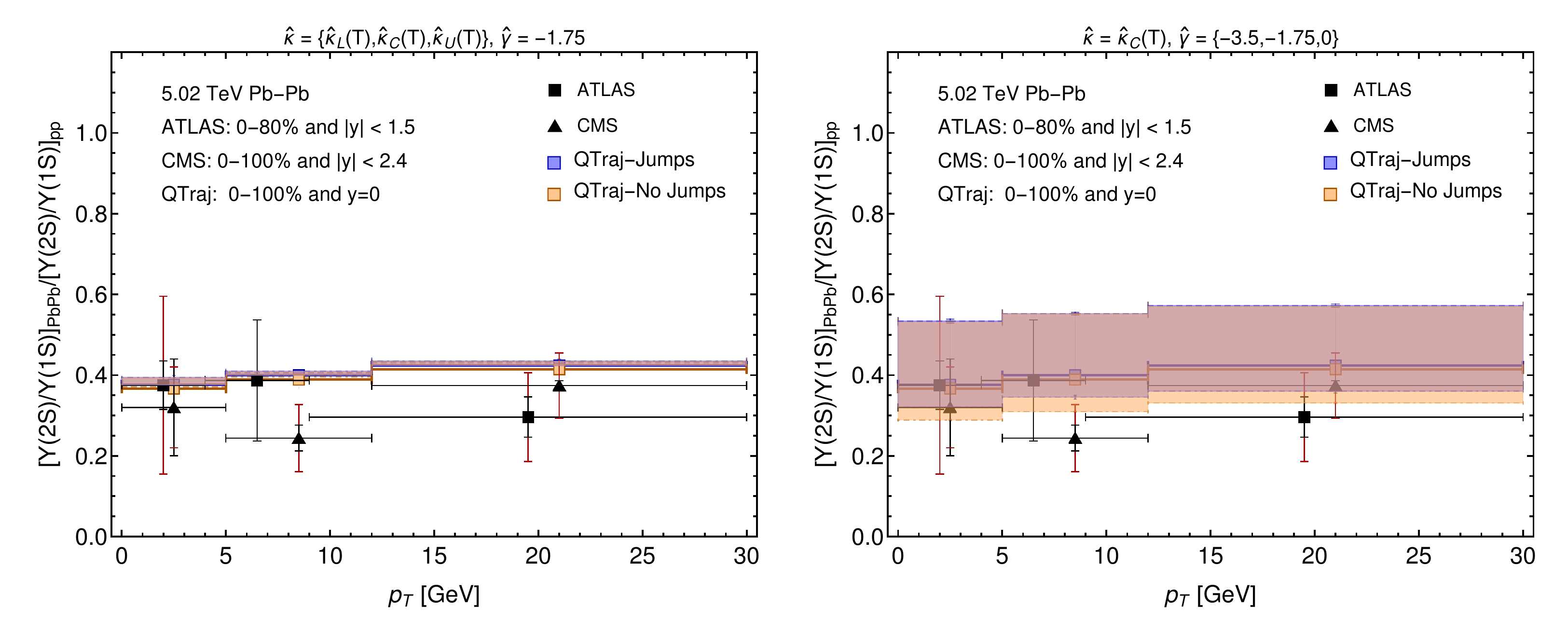}
	\end{center}
	\caption{(Color online)
		The double ratio of the nuclear modification factor $R_{AA}[\Upsilon(2S)]$ to $R_{AA}[\Upsilon(1S)]$ as a function of $p_{T}$ computed using \qtraj plotted against experimental measurements of $R_{AA}[\Upsilon(1S)]$ and $R_{AA}[\Upsilon(2S)]$ from the ATLAS~\cite{ATLAS5TeV}, and CMS~\cite{CMS:2017ycw} collaborations.
		We compare the results obtained using the full \qtraj algorithm (blue) with results obtained using evolution with $H_\text{eff}$ with no jumps (orange).
		The bands represent uncertainties as in Fig.~\ref{fig:raa_vs_npart}.
	}
	\label{fig:2S_double_ratio_vs_pt_jump_comp}
\end{figure*}

\begin{figure*}[t]
	\begin{center}
		\includegraphics[width=0.94\linewidth]{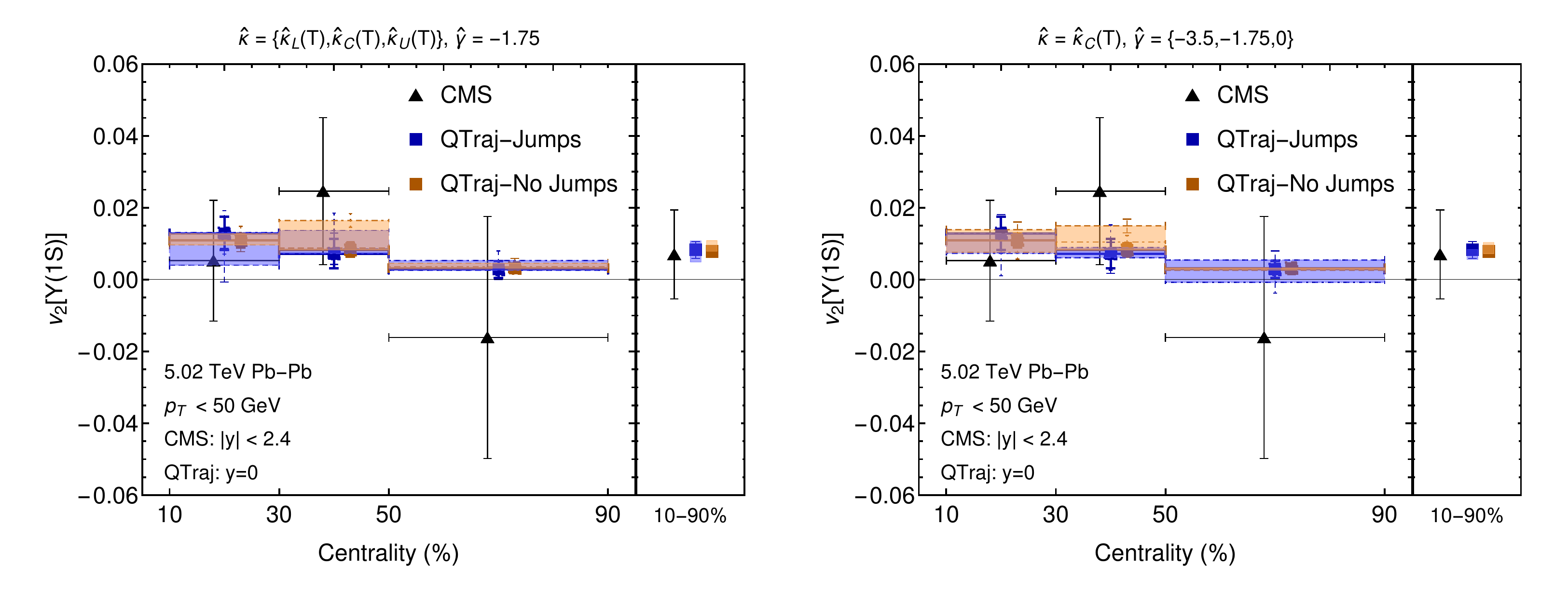}
	\end{center}
	\caption{(Color online)
		The elliptic flow $v_{2}$ of the $\Upsilon(1S)$ as a function of centrality computed using \qtraj plotted against experimental measurements of the CMS collaboration~\cite{CMS:2020efs}.
		We compare the results obtained using the full \qtraj algorithm (blue) with results obtained using evolution with $H_\text{eff}$ with no jumps (orange).
		The parameter variation is as in Fig.~\ref{fig:raa2S_vs_pt_jump_comp}.
	}
	\label{fig:v2_1S_vs_centrality_jump_comp}
\end{figure*}

\section{Comparisons of jump vs no jump evolution}
\label{app:jump_nojump}

In this appendix, we present comparisons between the full Lindblad evolution including the effects of quantum jumps and evolution in which we only evolve the system with the complex Hamiltonian $H_\text{eff}$.  This will help us to assess the role played by quantum jumps and their final effect on experimental observables.

In Fig.~\ref{fig:raa2S_vs_pt_jump_comp}, we plot a comparison of the \qtraj results for $R_{AA}[2S]$ as a function of $p_{T}$ implementing the full evolution with jumps to those obtained using only $H_\text{eff} $.
Each panel presents results obtained using different values of $\hat{\kappa}(T)$ and $\hat{\gamma}$.
We observe agreement to within uncertainties with the experimental data for all values of $\hat{\kappa}(T)$ and $\hat{\gamma}$ and between the full and $H_\text{eff} $ evolution for all values except $\hat{\gamma}=0$ (lower left panel). We note that the difference between the full Lindblad evolution and the $H_\text{eff}$ evolution is much smaller than the uncertainty obtained by varying $\hat{\kappa}$ and $\hat{\gamma}$. Therefore, until more precise determinations of $\kappa$ and $\gamma$ are available, the error made by ignoring jumps when computing $R_{AA}$ is negligible. 

In Fig.~\ref{fig:2S_double_ratio_vs_pt_jump_comp}, we present a comparison of results obtained using full evolution with jumps against results obtained using only $H_\text{eff} $ evolution for the double ratio $R_{AA}[\Upsilon(2S)]/R_{AA}[\Upsilon(1S)]$ as a function of $p_{T}$.
As in the case of $R_{AA}[\Upsilon(2S)]$, we observe the largest effect of the jumps in the case $\hat{\gamma}=0$ and $\hat{\kappa}=\hat{\kappa}_C$ (lower left panel in Fig.~\ref{fig:raa2S_vs_pt_jump_comp}) and in the case $\hat{\gamma}=-1.75$ and $\hat{\kappa}=\hat{\kappa}_{U}$ (upper right panel in Fig.~\ref{fig:raa2S_vs_pt_jump_comp}) with agreement to within the reported statistical uncertainties for the other values.
We note that the error induced by ignoring the jumps is of the order of the uncertainty obtained by varying $\kappa$ but much smaller than uncertainty obtained by varying $\gamma$. In summary, the uncertainty on the prediction of the double ratio $R_{AA}[\Upsilon(2S)]/R_{AA}[\Upsilon(1S)]$ as a function of $p_{T}$ is driven by $\gamma$ and a precise value of this quantity can potentially constrain the transport coefficient.

In Fig.~\ref{fig:v2_1S_vs_centrality_jump_comp}, we plot a comparison of full and $H_\text{eff} $ evolution results for $v_{2}[\Upsilon(1S)]$ as a function of $p_{T}$;
the panels correspond to separate variation of $\hat{\kappa}(T)$ (left panel) or $\hat{\gamma}$ (right panel), while the other parameter is kept fixed.
We observe again agreement to within uncertainties with the available experimental data. It is interesting to note that $v_2$ seems to be the only observable, within our obtained accuracy, in which the effect of the jumps competes with the uncertainties associated with the variation of $\kappa$ and $\gamma$. Therefore, $v_2$ appears to be the observable most sensitive to quantum jumps and might provide, in the future, an observable that cannot be explained with purely $H_\text{eff}$ evolution.

\newpage

\bibliography{qtraj2} 
\bibliographystyle{apsrev4-1}

\end{document}